 \documentclass[final,5p,times,twocolumn]{elsarticle}

\usepackage{amssymb}
\usepackage{amsmath}
\usepackage{amsthm}

\usepackage{lineno}
\usepackage{hyperref}

\usepackage[dvipsnames]{xcolor}
\definecolor{sPointColor}{HTML}{d300ff}
\definecolor{pPointColor}{HTML}{ff0081}
\usepackage{graphicx}
\usepackage{amsmath,bm,amsfonts,mathtools,amssymb}
\usepackage[toc,page]{appendix}
\usepackage{array}
\usepackage{paralist}
\usepackage{booktabs}
\usepackage{subcaption}
\usepackage{stfloats}

\usepackage{pgfplotstable}
\usepackage{pgfplots}
\pgfplotsset{compat=1.18}
\usetikzlibrary{plotmarks}
\usetikzlibrary{positioning}
\usetikzlibrary{matrix}
\usetikzlibrary{external}
\usetikzlibrary{calc}
\usetikzlibrary{arrows}
\usetikzlibrary{decorations.markings}
\usetikzlibrary{shapes.geometric}
\usepgfplotslibrary{fillbetween}
\usepgfplotslibrary{groupplots}

\pgfplotsset{select coords between index/.style 2 args={
    x filter/.code={
        \ifnum\coordindex<#1\fi
        \ifnum\coordindex>#2\fi
    }
}}

\usepackage{algorithm}
\usepackage{algpseudocode}
\usepackage{booktabs}
\usepackage{array}



\journal{Computer Physics Communications}

\begin{document}

\begin{frontmatter}

\title{A Fast Spectral Formulation of the Multiscale Proper Orthogonal Decomposition} 

\author[ctu,itcas,vki]{Marek Belda} 
\author[vki,vub]{Lorenzo Schena} 
\author[vki]{Romain Poletti} 
\author[itcas]{Martin Isoz}
\author[ctu]{Tom{áš} Hyhl{í}k}
\author[vki,atm,aerg]{Miguel Alfonso Mendez\corref{cor1}} 
\ead{mendez@vki.ac.be} 

\affiliation[ctu]{organization={Department of Fluid Dynamics and Thermodynamics, Faculty of Mechanical Engineering, Czech Technical University in Prague},
            addressline={Technick{á}~1902/4}, 
            city={Prague~6 -- Dejvice},
            postcode={160~00}, 
            country={Czech Republic}}

\affiliation[itcas]{organization={Institute of Thermomechanics, Czech Academy of Sciences},
            addressline={Dolej{š}kova~1402/5}, 
            city={Prague~8 -- Libe{ň}},
            postcode={182~00}, 
            country={Czech Republic}}
            
\affiliation[vki]{organization={The von Karman Institute for Fluid Dynamics},
            addressline={Waterloosesteenweg 72}, 
            city={Sint-Genesius-Rode},
            postcode={B-1640}, 
            country={Belgium}}    

\affiliation[vub]{organization={Vrije Universiteit Brussel (VUB)},
            addressline={Pleinlaan 2, Elsene}, 
            city={Brussel},
            postcode={1050}, 
            country={Belgium}}   
            
\affiliation[atm]{organization={Aero-Thermo-Mechanics Laboratory, École Polytechnique de Bruxelles, Université Libre de Bruxelles},
            addressline={Av. Franklin Roosevelt 50}, 
            city={Brussels},
            postcode={1050}, 
            country={Belgium}} 

\affiliation[aerg]{organization={Aerospace Engineering Research Group, Universidad Carlos III de Madrid},
            addressline={Av. de la Universidad 30}, 
            city={Leganés},
            postcode={28911}, 
            country={Spain}} 

\begin{abstract}

Multiscale Proper Orthogonal Decomposition (mPOD) decomposes fluid flows into energy-optimal modes within prescribed frequency bands by combining Proper Orthogonal Decomposition with a multiresolution analysis (MRA). In its classical formulation, mPOD relies on a filter bank of finite impulse response (FIR) filters, enabling lossless reconstruction while mitigating Gibbs oscillations and temporal ringing. However, the smooth transition bands required for this purpose introduce partial spectral overlap between adjacent scales and require, for each band, the solution of an eigenvalue problem spanning the full temporal dimension.

This work introduces a fast spectral formulation of the mPOD that substantially reduces the computational cost. The proposed approach replaces time-domain FIR filters with compact spectral masks enforcing strictly disjoint frequency supports, thereby exactly decoupling the problem across scales. This leads to a block-diagonal correlation operator in spectral space, so that each band can be treated independently. The resulting eigenvalue problems reduce to small systems whose size depends on the number of active frequencies per band rather than the full time dimension.  The approach is validated on a synthetic dataset highlighting spectral windowing effects and on experimental particle image velocimetry (PIV) data of a cylinder wake at Reynolds number \(\mbox{Re} \approx 5000\). In both cases, the proposed formulation accurately recovers the modal structures and singular values of the classical mPOD while reducing the computational cost by orders of magnitude.

\end{abstract}




\vfill



\end{frontmatter}


\section{Introduction}
\label{sec:intro}
Data-driven modal decompositions are key tools for identifying coherent structures and interpretable patterns in fluid flow data, enabling diagnostics and reduced-order modeling~\citep{Taira2017,Begiashvili2023,Mendez_2023}.

A taxonomy of such methods can be organized along a spectrum between two competing objectives: energy optimality, i.e., the ability to capture the largest fraction of the dataset's variance with a truncated basis, and spectral purity, i.e., the localization of each mode within a prescribed frequency band. At these two extremes sit POD~\citep{Lumley1970,sirovich1987,Dawson_2023} and DMD~\citep{schmid2010,Tu2014jcd}, respectively. The classical trade-off between the two has motivated the development of hybrid decompositions such as the spectral POD proposed by \citet{Sieber2016}, the frequency-based formulations of the POD by \citet{Towne2018}, independently referred to as spectral POD, and the multiscale POD (mPOD) by \citet{Mendez2019}.

Unlike other hybrid decompositions, the mPOD gives the user direct control over this trade-off by explicitly partitioning the frequency axis into prescribed bands and providing an energy-optimal decomposition within each. This yields modes that are simultaneously energy-ranked and spectrally constrained, enabling consistent separation of multiscale and transient dynamics while retaining the convergence properties of classical POD. This flexibility made the mPOD relevant for a wide range of problems, including turbulent motion analysis~\citep{Chi2022}, controlled cylinder wakes~\citep{Zhong2023}, transient dynamics \cite{Mendez2020}, jet wiping and gas--liquid coupling instabilities~\citep{barreiro-villaverde_coupling_2024}, cavitating orifice flows~\citep{Esposito2021}, propeller wakes~\citep{Huang2025}, swirled and stratified flames~\citep{Procacci2022}, swirling spray flames near lean blowout~\citep{DeGiorgi2024}, flow-induced vibration problems~\citep{Janocha2022}, compressor corner separation~\citep{Liu2025}, spray characterization~\citep{Zhou2022}, sweeping-jet flow control~\citep{Wang2026} or flows over low-aspect-ratio plates~\citep{Zhu2025} to mention some examples.

The spectral constraints of mPOD were first defined via wavelet-based multiresolution analysis~\citep{Mendez2018} and later reformulated using filter banks~\citep{Mendez2019,Mendez_2023}, as implemented in the open-source toolbox \textsc{MODULO}~\citep{Ninni2020,Poletti2024}. In this approach, scale separation is achieved through time-domain FIR filters with smooth transition bands, minimizing spectral leakage and temporal ringing but precluding strict separation between adjacent bands and requiring a full eigenvalue problem per band, sized by the number of temporal snapshots rather than band content. This cost scales steeply with dataset size and dominates the computation for large problems.

Rapid growth in dataset size has spurred efficient implementations of data decomposition methods, including randomized SVD~\citep{palitta2025}, streaming spectral POD~\citep{Schmidt2019}, and tensor-accelerated DMD variants~\citep{Li2023,he2025}. In this context, we propose a fast spectral variant of multiscale Proper Orthogonal Decomposition (mPOD) that greatly reduces computational cost.

Our approach departs from the classical formulation by strictly separating spectral supports. While the idea of tackling the eigenvalue problem in the frequency domain can be traced back to the early spectral formulations of the POD~\citep{George1988,Glauser1987}, the present compact formulation, restricted to non-zero spectral content, is, to the authors’ knowledge, the first systematic framework for multiscale decomposition of this kind.

The remainder of this paper is organized as follows. Section\ref{sec2} briefly recalls the classical formulation of the mPOD and highlights its computational bottlenecks. Section~\ref{sec3} introduces the proposed spectral formulation, detailing the construction of the compact operators and the resulting fast algorithm, proposed in two variants. Section \ref{sec4} verifies the approach on both synthetic and experimental datasets and assesses its computational performance compared to the classical mPOD. Finally, concluding remarks and perspectives are provided in Section \ref{sec5}.
\section{The classical Multiscale POD (mPOD)}
\label{sec2}

We denote $\mathbf{D}\in\mathbb{R}^{n_s\times n_t}$ the snapshot matrix collecting $n_t$ temporal realizations of a spatial field sampled at $n_s$ spatial locations. The mPOD proceeds in three steps: (i) the snapshot matrix is partitioned into scale-specific contributions $\mathbf{D}_m$ using a predefined filter bank, (ii) a POD is performed on each scale independently, and (iii) the resulting modes are assembled into a single global factorization. This yields a decomposition of the form
\begin{equation}
\label{eq1}
\mathbf{D}=\sum_{m=1}^{n_M}\mathbf{D}_m
= \sum_{m=1}^{n_M}\mathbf{\Phi}_m\mathbf{\Sigma}_m\mathbf{\Psi}_m^{\top}
= \mathbf{\Phi}_\mathcal{M}\mathbf{\Sigma}_\mathcal{M}\mathbf{\Psi}_\mathcal{M}^{\top}\;,
\end{equation}
where $(\cdot)^{\top}$ denotes transposition, $\mathbf{\Psi}_m\in\mathbb{R}^{n_t \times r_m}$ contains the $r_m$ temporal modes associated with scale $m$, $\mathbf{\Phi}_m \in \mathbb{R}^{n_s \times r_m}$ the corresponding spatial modes, and $\mathbf{\Sigma}_m \in \mathbb{R}^{r_m \times r_m}$ is a diagonal matrix of singular values. The temporal basis of the mPOD is obtained by concatenating the temporal structures from all scales,
\begin{equation}
\mathbf{\Psi}_{\mathcal{M}}=
\left[
\mathbf{\Psi}_1,\mathbf{\Psi}_2,\dots,\mathbf{\Psi}_{n_M} 
\right]\in\mathbb{R}^{n_t \times R}\;,
\end{equation} where $R=\sum_{m=1}^{n_M} r_m$. The associated spatial structures $\mathbf{\Phi}_\mathcal{M}$ and amplitudes $\mathbf{\Sigma}_\mathcal{M}$ are then obtained by projection and normalization, as in the standard snapshot POD \cite{sirovich1987} (see also \citet{Mendez_2023}).

The partition of the snapshot matrix into scale contributions $\mathbf{D}_m$ defines a multi-resolution analysis of the snapshot matrix (see \cite{Strang1997}). Each scale is identified by a temporal filter. In the time domain, the filtering corresponds to a convolution along the temporal direction,
\[
\mathbf{D}_m=\mathbf{D}\mathbf{H}_m ,
\]
where $\mathbf{H}_m\in\mathbb{R}^{n_t\times n_t}$ collects the impulse responses of the filter associated with scale $m$ (see \cite{Mendez_2023}). 

Equivalently, filtering can be performed in the frequency domain. Denoting by $\bm{\Psi}_\mathrm{F}\in\mathbb{C}^{n_t\times n_t}$ the discrete Fourier matrix for the temporal direction, such that the temporal Fourier coefficients of the data read
\(
\widehat{\mathbf{D}}=\mathbf{D}\,\overline{\bm{\Psi}}_\mathrm{F},
\)
with $\overline{\bm{\Psi}}_\mathrm{F}$ denoting a complex conjugation of $\bm{\Psi}_\mathrm{F}$, and denoting by $\widehat{\mathbf{H}}_m$ the diagonal matrix collecting the transfer function of the filter associated with scale $m$, the scale contribution reads
\begin{equation}
\label{eq_Dm}
\mathbf{D}_m
=
\mathbf{D}\,\overline{\bm{\Psi}}_\mathrm{F}\,
\widehat{\mathbf{H}}_m\,
\bm{\Psi}_\mathrm{F}.
\end{equation}

In \eqref{eq_Dm} and in the following, matrix multiplications involving $\bm{\Psi}_\mathrm{F}$ and $\overline{\bm{\Psi}}_\mathrm{F}$ are only formal-- in practice, these operations correspond to inverse and forward discrete Fourier transforms implemented using \texttt{fft} and \texttt{ifft} routines.

Assuming perfect spectral separation between filters, that is, 
\begin{equation}
\label{PUM}
\sum_{m=1}^{n_M}\widehat{\mathbf{H}}_m = \mathbf{I},
\qquad
\widehat{\mathbf{H}}_m\,\widehat{\mathbf{H}}_\ell = \mathbf{0}, \quad m\neq\ell,
\end{equation}
each frequency is assigned to a single scale.

Therefore, each matrix $\mathbf{D}_m$ has rank $r_m \le n^{(m)}$, where $n^{(m)}$ denotes the number of Fourier coefficients retained in the $m$-th scale. 
For real-valued data, Fourier coefficients occur in complex-conjugate pairs, so the number of independent degrees of freedom is generally smaller than $n^{(m)}$.

 If \eqref{PUM} holds, the temporal modes belonging to different scales are mutually orthogonal and $\mathbf{\Psi}_{\mathcal{M}}$ forms an orthonormal temporal basis-- the factorization \eqref{eq1} is a summation of Singular Value Decompositions (SVDs) of the filtered blocks. Accordingly, the spatial modes $\mathbf{\Phi}_m$ are orthonormal within each scale by construction of the SVD, but are not, in general, orthogonal across different scales.

In practice, preserving the orthogonality of the temporal basis $\mathbf{\Psi}_{\mathcal M}$ requires additional processing. The assumption of perfect spectral separation is not enforced because filters with excessively sharp transfer functions would produce long impulse responses and spurious oscillations in the time domain. For this reason, practical filter banks employ smooth transition bands, which lead to partial spectral overlap between neighboring filters. To restore orthogonality, the traditional mPOD algorithm uses a QR factorization \(
\mathbf{\Psi}_{\mathcal M}=\mathbf{Q}\mathbf{R},
\)to polish the assembled temporal basis, i.e. replacing $\mathbf{\Psi}_{\mathcal M}\leftarrow\mathbf{Q}$. 

\subsection{Case $n_t \ll n_s$ -- temporal correlation approach}

When the number of temporal samples is much smaller than the number of spatial degrees of freedom, it is convenient to construct the decomposition from the temporal correlation matrix
\(
\mathbf{K}=\mathbf{D}^{\dag}\mathbf{D}\in\mathbb{R}^{n_t\times n_t}
\), where $\dag$ denotes the Hermitian transpose. This leads to eigenvalue problems of size $n_t\times n_t$.

The multiscale decomposition is obtained by neglecting cross terms between scales (which should equal zero in case of perfect spectral separation). That is
\begin{equation}
\begin{aligned}
\mathbf{K}
&=\mathbf{D}^{\dag}\mathbf{D}
=\left(\sum_{m=1}^{n_M}\mathbf{D}_m\right)^{\dag}
\left(\sum_{m=1}^{n_M}\mathbf{D}_m\right) \\
&=\sum_{m=1}^{n_M}\mathbf{D}_m^{\dag}\mathbf{D}_m
+\sum_{\ell\neq m}\mathbf{D}_\ell^{\dag}\mathbf{D}_m
\;\approx\;
\sum_{m=1}^{n_M}\mathbf{D}_m^{\dag}\mathbf{D}_m = \sum^{n_M}_{m=1} \mathbf{K}_m.
\end{aligned}
\end{equation}

Introducing \eqref{eq_Dm}, one obtains
\begin{equation}
\label{K_m}
\begin{aligned}
\mathbf{K}_m
&=
\left(
\mathbf{D}\,\overline{\bm{\Psi}}_{\mathrm F}\,
\widehat{\mathbf H}_m\,
\bm{\Psi}_{\mathrm F}
\right)^{\dag}
\left(
\mathbf{D}\,\overline{\bm{\Psi}}_{\mathrm F}\,
\widehat{\mathbf H}_m\,
\bm{\Psi}_{\mathrm F}
\right) \\
&=
\overline{\bm{\Psi}}_{\mathrm F}\,
\widehat{\mathbf H}_m\,
{\bm{\Psi}}_{\mathrm F}
\mathbf{K}\,
\overline{\bm{\Psi}}_{\mathrm F}\,
\widehat{\mathbf H}_m\,
\bm{\Psi}_{\mathrm F}\\
&=
\overline{\bm{\Psi}}_{\mathrm F}\,
\widehat{\mathbf H}_m\,
\mathbf{K}_\mathcal{F}\,
\widehat{\mathbf H}_m\,
\bm{\Psi}_{\mathrm F}\,\\
&=
\overline{\bm{\Psi}}_{\mathrm F}\,
\mathbf{K}_{\mathcal{F},m}\,
\bm{\Psi}_{\mathrm F}\,,
\end{aligned}
\end{equation}
assuming real-valued transfer functions (corresponding to non-causal, zero-phase filtering), and having introduced the spectral representation of the temporal correlation matrix
\begin{equation}
\label{K_F}
\mathbf{K}_{\mathcal{F}}
=
{\bm{\Psi}}_{\mathrm F}\,
\mathbf{K}\,
\overline{\bm{\Psi}}_{\mathrm F}\,,
\end{equation} and its filtered version $\mathbf{K}_{\mathcal{F},m}=\widehat{\mathbf H}_m\,\mathbf{K}_{\mathcal F}\,\widehat{\mathbf H}_m$.

These spectral representations correspond to applying the discrete Fourier transform along both temporal indices of the correlation matrix (see \cite{Mendez2019}).

The temporal structures are eigenvectors of the covariance matrix at each scale, that is 
\begin{equation}
\label{K_eigen}
\mathbf{K}_m\,\mathbf{\Psi}_m= \mathbf{\Psi}_m \mathbf{\Sigma}^2_m \,.
\end{equation} 

Algorithm \ref{alg:classicK} summarizes the classic mPOD algorithm proposed in \cite{Mendez2019}. To set the stage for the proposed spectral implementation, it is worth noticing that the definition in \eqref{K_F} leads to a similarity transform $
\mathbf{K}_{\mathcal{F},m}=\mathbf{\Psi}_\mathrm{F} \mathbf{K}_m\,\overline{\mathbf{\Psi}}_\mathrm{F}$ and hence the eigenvalue problem \eqref{K_eigen} can be related to the eigenvalue problem for $\mathbf{K}_{\mathcal{F},m}$ \cite{Mendez_2023}:

\begin{equation}
\label{K_F_eig}
\mathbf{K}_{\mathcal{F},m}\,\widehat{\mathbf{\Psi}}_m
=
\widehat{\mathbf{\Psi}}_m \mathbf{\Sigma}_m^2,
\end{equation}
where $\widehat{\mathbf{\Psi}}_m=\bm{\Psi}_{\mathrm F}\mathbf{\Psi}_m$ is the Fourier representation of the temporal modes. The modes in physical space are recovered via
\begin{equation}
\label{Psi_M_F}
\mathbf{\Psi}_m
=
\overline{\bm{\Psi}}_{\mathrm F}\,
\widehat{\mathbf{\Psi}}_m.
\end{equation}

Equations \eqref{K_F_eig} and \eqref{Psi_M_F} could be used instead of \eqref{K_eigen} in line 6 of Algorithm~\ref{alg:classicK}. However, since both involve diagonalization of matrices of size $n_t\times n_t$, there are currently no benefits in operating in the spectral domain.

\begin{algorithm}[htb]
\caption{Classical mPOD via temporal-correlation}
\label{alg:classicK}
\begin{algorithmic}[1]
\State \textbf{Input:} data matrix $\mathbf D$, frequency splitting vector $F_V$
\State Construct the filter bank $\{\widehat{\mathbf H}_m\}_{m=1}^{n_M}$ from $F_V$
\State Compute temporal correlation matrix
\(
\mathbf K=\mathbf D^\dag \mathbf D
\)
\For{$m=1,\dots,n_M$}
\State Compute filtered correlation matrix
\(
\mathbf K_m
=
\mathbf H_m^\dagger \mathbf K \mathbf H_m
\)
\State Solve eigenvalue problem
\(
\mathbf K_m \mathbf \Psi_m
=
\mathbf \Psi_m \mathbf \Sigma_m^2
\)
\EndFor
\State Assemble temporal basis
\(
\mathbf \Psi_{\mathcal M}^0
=
[\mathbf \Psi_1,\dots,\mathbf \Psi_{n_M}]
\)
\State Enforce orthonormality
\(
\mathbf \Psi_{\mathcal M}^0
=
\mathbf Q\mathbf R,
\qquad
\mathbf \Psi_{\mathcal M}=\mathbf Q
\)
\State Compute spatial modes
\(
\mathbf \Phi_{\mathcal M}
=
\mathbf D\,\mathbf \Psi_{\mathcal M}\,
\mathbf \Sigma_{\mathcal M}^{-1}
\)
\end{algorithmic}
\end{algorithm}

The computational cost is dominated by two operations: the construction of the temporal correlation matrix $\mathbf K=\mathbf D^\dag \mathbf D$, which requires
\(O(n_s n_t^2)\) operations, and the repeated solution of the eigenvalue problems of size $n_t\times n_t$, which requires \(O(n_t^3)\) per scale. Therefore, the overall complexity is
\begin{equation}
\label{eq:FIRCompl}
O\!\left(
n_s n_t^2 + n_M n_t^3
\right).
\end{equation}
 
\subsection{Case $n_s \ll n_t$--  spatial correlation approach}

When the number of spatial degrees of freedom is much smaller than the number of temporal samples, it is more convenient to construct the decomposition directly from the matrices $\mathbf{D}_m$ in \eqref{eq_Dm} through their singular value decomposition. 

This can be computed by diagonalizing the spatial correlation matrices $\mathbf{C}_m=\mathbf{D}_m\mathbf{D}_m^\dag \in \mathbb{R}^{n_s\times n_s}$, which leads to $n_M$ eigenvalue problems of size $n_s\times n_s$. An efficient spectral formulation to this approach, not presented in \cite{Mendez2019}, is detailed below. 

Introducing \eqref{eq_Dm}, one obtains
\begin{equation}
\label{eqCm}
\begin{aligned}
\mathbf C_m
&=
\left(
\mathbf D\,\overline{\mathbf \Psi}_{\mathrm F}\,
\widehat{\mathbf H}_m\,
\mathbf \Psi_{\mathrm F}
\right)
\left(
\mathbf D\,\overline{\mathbf \Psi}_{\mathrm F}\,
\widehat{\mathbf H}_m\,
\mathbf \Psi_{\mathrm F}
\right)^\dagger \\
&=
\mathbf D\,\overline{\mathbf \Psi}_{\mathrm F}\,
\widehat{\mathbf H}_m\,
\mathbf \Psi_{\mathrm F}
\mathbf \Psi_{\mathrm F}^\dagger
\widehat{\mathbf H}_m^\dagger
\mathbf \Psi_{\mathrm F}
\mathbf D^\dagger \\
&=
\mathbf D\,\overline{\mathbf \Psi}_{\mathrm F}\,
\widehat{\mathbf H}_m^2\,
\mathbf \Psi_{\mathrm F}
\mathbf D^\dagger \\
&=
\widehat{\mathbf D}\,
\widehat{\mathbf H}_m^2\,
\widehat{\mathbf D}^\dagger\\
&=
\widehat{\mathbf D}_{m}\widehat{\mathbf D}_{m}^\dagger .
\end{aligned}
\end{equation}

Equation \eqref{eqCm} shows that the spatial correlation matrix at scale $m$ can be assembled directly in the spectral domain, without explicitly forming the filtered signal in the time domain. Once the eigenvalue problem
\begin{equation}
\mathbf C_m \mathbf \Phi_m
=
\mathbf \Phi_m \mathbf \Sigma_m^2
\end{equation}
has been solved, the temporal modes can be recovered from the filtered spectral data through
\begin{equation}
\widehat{\mathbf \Psi}_m
=
\widehat{\mathbf D}_{m}^\dagger
\mathbf \Phi_m
\mathbf \Sigma_m^{-1},
\qquad
\mathbf \Psi_m
=
\overline{\mathbf \Psi}_{\mathrm F}\,
\widehat{\mathbf \Psi}_m .
\end{equation}

Algorithm~\ref{alg:classicC} summarizes the corresponding mPOD implementation.

\begin{algorithm}[htb]
\caption{Classical mPOD via spatial correlation}
\label{alg:classicC}
\begin{algorithmic}[1]
\State \textbf{Input:} data matrix $\mathbf D$, frequency splitting vector $F_V$
\State Construct the filter bank $\{\widehat{\mathbf H}_m\}_{m=1}^{n_M}$ from $F_V$.
\State Compute the Fourier-transformed data matrix
\(
\widehat{\mathbf D}
=
\mathbf D\,\overline{\mathbf \Psi}_{\mathrm F}
\)
\For{$m=1,\dots,n_M$}
\State Filter the data and get
\(
\widehat{\mathbf D}_{m}
=
\widehat{\mathbf D} \widehat{\mathbf H}_m
\)
\State Compute the spatial correlation matrix
\(
\mathbf C_m
=
\widehat{\mathbf D}_{m}\widehat{\mathbf D}_{m}^\dagger
\)
\State Solve eigenvalue problem
\(
\mathbf C_m \mathbf \Phi_m
=
\mathbf \Phi_m \mathbf \Sigma_m^2
\)
\State Compute spectra \(
\widehat{\mathbf \Psi}_m
=
\widehat{\mathbf D}_{m}^\dagger
\mathbf \Phi_m
\mathbf \Sigma_m^{-1}
\)
\State Recover temporal modes
\(
\mathbf \Psi_m
=
\overline{\mathbf \Psi}_{\mathrm F}\widehat{\mathbf \Psi}_m
\)
\EndFor
\State Assemble temporal basis
\(
\mathbf \Psi_{\mathcal M}^0
=
[\mathbf \Psi_1,\dots,\mathbf \Psi_{n_M}]
\)
\State Enforce orthonormality
\(
\mathbf \Psi_{\mathcal M}^0
=
\mathbf Q\mathbf R,
\qquad
\mathbf \Psi_{\mathcal M}=\mathbf Q
\)
\State Compute spatial modes
\(
\mathbf \Phi_{\mathcal M}
=
\mathbf D\,\mathbf \Psi_{\mathcal M}\,
\mathbf \Sigma_{\mathcal M}^{-1}
\)
\end{algorithmic}
\end{algorithm}

The computational cost of this formulation is dominated by three operations: the Fourier transform of the data matrix, which is \(O(n_s n_t \log n_t)\), the construction of the filtered spatial correlation matrices, which is $O\!\left(n_s^2 n_t\right)$ for each frequency band, and the repeated solution of the eigenvalue problems of size $n_s\times n_s$, which is \(O(n_M n_s^3)\). The total complexity of this approach is thus
\begin{equation}
O\!\left(
n_s n_t \log n_t
+
n_M n_s^2 n_t
+
n_M n_s^3
\right).
\end{equation}

\section{The fast (spectral) mPOD }\label{sec3}

The classical mPOD employs smooth transfer functions $H_m(f)$ designed to approximate a partition of unity while limiting Gibbs oscillations and temporal ringing. However, the resulting overlap between neighboring bands prevents strictly compact spectral support, so the matrices $\mathbf{D}_m$ and $\mathbf{K}_m$ are not exactly low-rank, even if most of their energy is concentrated in a subspace of dimension close to $n^{(m)}$.

In the present work, we deliberately adopt the opposite design philosophy in order to unlock major computational gains for large datasets. Instead of constructing time-domain FIR filters, we define compact spectral masks $M_m(f)$ directly in the frequency domain, enforcing strictly disjoint spectral supports while relaxing the partition-of-unity condition.

Let $F_V=\{f_1,\dots,f_{n_M-1}\}$ denote the frequency splitting vector defining the spectral bands. For each band $m$, the mask $M_m(f)$ is constructed to be unity within the core of the interval $[f_{m-1},f_m]$ and to smoothly taper to zero near the band boundaries. The tapering is implemented through cosine windows of width $\delta_m$ placed inside the nominal band, so that
\(
M_m(f)=0 \quad \text{for } |f|<f_{m-1} \text{ and } |f|>f_m ,
\)
while smoothly reaching unity within the band interior.

Because the transition regions are contained within each band rather than extending across adjacent bands, neighboring masks share exact zero crossings at the splitting frequencies. Therefore,
\(
M_\ell(f)\,M_m(f)=0\), if \( \ell\neq m\),
and the filtered matrices $\mathbf{D}_m$ and $\mathbf{K}_m$ contain at most $n^{(m)}$ active spectral components. Hence,
\(
\mathrm{rank}(\mathbf{D}_m),\ \mathrm{rank}(\mathbf{K}_m)\le n^{(m)}.
\) This compact-support property enables a reformulation of the mPOD in which each scale can be treated independently on a reduced spectral subspace. The structure of this reduction is made explicit by examining the sparsity patterns of the filtered spectral data matrices $\widehat{\mathbf D}_m$ and their associated correlation matrices $\mathbf K_{\mathcal F,m}$.

\subsection{Sparsity patterns and compact structures}
The sparsity patterns in $\widehat{\mathbf D}_m$ and $\mathbf K_{\mathcal F,m}$ depend on the position of the passband within the spectrum and fall into three canonical configurations. Since $\mathbf D\in\mathbb{R}^{n_s\times n_t}$ and $\mathbf K=\mathbf D^\dag \mathbf D\in\mathbb{R}^{n_t\times n_t}$ are real-valued, their Fourier representations are conjugate symmetric with respect to positive and negative frequencies. As a result, the spectral blocks associated with opposite frequency intervals are not independent, but related by complex conjugation (up to a possible permutation of the frequency ordering).

For the first scale, corresponding to the low-frequency contribution, only the coefficients in a neighborhood of zero frequency are retained. In a frequency ordering centred around $f=0$, the filtered spectral data matrix takes the form
\begin{equation}
\widehat{\mathbf D}_1 =
\begin{bmatrix}
\mathbf 0 & \widehat{\mathbf D}_{\mathcal L} & \mathbf 0
\end{bmatrix},
\end{equation}
where $\widehat{\mathbf D}_{\mathcal L}\in\mathbb{C}^{n_s\times n^{(1)}}$ collects the active low-frequency coefficients and the zero blocks represent frequency intervals where the spectral mask vanishes. The size of these blocks is determined by the bandwidth of the active region, so that each matrix $\widehat{\mathbf D}_m$ contains at most $n^{(m)}$ nonzero columns (up to possible degeneracies in the data) corresponding to the retained frequencies, while the remaining $n_t - n^{(m)}$ columns are identically zero.

The corresponding filtered spectral correlation matrix is
\begin{equation}
\mathbf K_{\mathcal F,1} =
\begin{bmatrix}
\mathbf 0 & \mathbf 0 & \mathbf 0 \\
\mathbf 0 & \mathbf K_{\mathcal F,\mathcal L} & \mathbf 0 \\
\mathbf 0 & \mathbf 0 & \mathbf 0
\end{bmatrix},
\end{equation}
with $\mathbf K_{\mathcal F,\mathcal L}=\widehat{\mathbf D}_{\mathcal L}^\dagger \widehat{\mathbf D}_{\mathcal L}$ the compact version of the large scale contribution.

For an intermediate scale $m=2,\dots,n_M-1$, corresponding to a band-pass contribution, the retained coefficients occupy two symmetric intervals around the zero frequency. The filtered spectral data matrix takes the form
\begin{equation}
\widehat{\mathbf D}_m =
\begin{bmatrix}
\mathbf 0 & \widehat{\mathbf D}_{\mathcal B}^{(-)} & \mathbf 0 & \widehat{\mathbf D}_{\mathcal B}^{(+)} & \mathbf 0
\end{bmatrix},
\end{equation}
where $\widehat{\mathbf D}_{\mathcal B}^{(-)}$ and $\widehat{\mathbf D}_{\mathcal B}^{(+)}$ correspond to the negative- and positive-frequency contributions, respectively, and satisfy
\begin{equation}
\widehat{\mathbf D}_{\mathcal B}^{(+)}
=
\overline{\widehat{\mathbf D}_{\mathcal B}^{(-)}}
\qquad
\text{(up to column ordering).}
\end{equation}
The filtered spectral correlation matrix then has the structure
\begin{equation}
\mathbf K_{\mathcal F,m} =
\begin{bmatrix}
\mathbf 0 & \mathbf 0 & \mathbf 0 & \mathbf 0 & \mathbf 0 \\
\mathbf 0 & \mathbf K_{\mathcal F,\mathcal B}^{(1)} & \mathbf 0 & \mathbf K_{\mathcal F,\mathcal B}^{(2)} & \mathbf 0 \\
\mathbf 0 & \mathbf 0 & \mathbf 0 & \mathbf 0 & \mathbf 0 \\
\mathbf 0 & \mathbf K_{\mathcal F,\mathcal B}^{(2)\dagger} & \mathbf 0 & \mathbf K_{\mathcal F,\mathcal B}^{(3)} & \mathbf 0 \\
\mathbf 0 & \mathbf 0 & \mathbf 0 & \mathbf 0 & \mathbf 0
\end{bmatrix},
\end{equation}
with
\begin{equation}
\label{compact_B}
\mathbf K_{\mathcal F,\mathcal B} =
\begin{bmatrix}
\mathbf K_{\mathcal F,\mathcal B}^{(1)} & \mathbf K_{\mathcal F,\mathcal B}^{(2)} \\
\mathbf K_{\mathcal F,\mathcal B}^{(2)\dagger} & \mathbf K_{\mathcal F,\mathcal B}^{(3)}
\end{bmatrix}
=
\widehat{\mathbf D}_{\mathcal B}^\dagger \widehat{\mathbf D}_{\mathcal B}
\end{equation} the compact version.
Moreover, because of the conjugate symmetry of the Fourier coefficients of real-valued signals, the diagonal blocks $\mathbf K_{\mathcal F,\mathcal B}^{(1)}$ and $\mathbf K_{\mathcal F,\mathcal B}^{(3)}$ are themselves related by complex conjugation, up to a possible permutation induced by frequency ordering. Finally, for the last scale, corresponding to the high-pass contribution, the retained coefficients occupy the outer portions of the spectrum. The filtered spectral data matrix takes the form
\begin{equation}
\widehat{\mathbf D}_{n_M} =
\begin{bmatrix}
\widehat{\mathbf D}_{\mathcal H}^{(-)} & \mathbf 0 & \widehat{\mathbf D}_{\mathcal H}^{(+)}
\end{bmatrix},
\end{equation}
where $\widehat{\mathbf D}_{\mathcal H}^{(-)}$ and $\widehat{\mathbf D}_{\mathcal H}^{(+)}$ denote the negative- and positive-frequency contributions, respectively, and satisfy
\begin{equation}
\widehat{\mathbf D}_{\mathcal H}^{(+)} =
\overline{\widehat{\mathbf D}_{\mathcal H}^{(-)}}
\qquad
\text{(up to column ordering).}
\end{equation}
The filtered spectral correlation matrix is
\begin{equation}
\mathbf K_{\mathcal F,n_M} =
\begin{bmatrix}
\mathbf K_{\mathcal F,\mathcal H}^{(1)} & \mathbf 0 & \mathbf K_{\mathcal F,\mathcal H}^{(2)} \\
\mathbf 0 & \mathbf 0 & \mathbf 0 \\
\mathbf K_{\mathcal F,\mathcal H}^{(2)\dagger} & \mathbf 0 & \mathbf K_{\mathcal F,\mathcal H}^{(3)}
\end{bmatrix},
\end{equation}
with
\begin{equation}
\label{compact_H}
\mathbf K_{\mathcal F,\mathcal H} =
\begin{bmatrix}
\mathbf K_{\mathcal F,\mathcal H}^{(1)} & \mathbf K_{\mathcal F,\mathcal H}^{(2)} \\
\mathbf K_{\mathcal F,\mathcal H}^{(2)\dagger} & \mathbf K_{\mathcal F,\mathcal H}^{(3)}
\end{bmatrix}
=
\widehat{\mathbf D}_{\mathcal H}^\dagger \widehat{\mathbf D}_{\mathcal H}
,
\end{equation}
 recalling that the conjugate symmetry of the Fourier coefficients of real-valued signals ensures that the diagonal blocks $\mathbf K_{\mathcal F,\mathcal H}^{(1)}$ and $\mathbf K_{\mathcal F,\mathcal H}^{(3)}$ are related by complex conjugation, up to a permutation induced by frequency ordering.

\subsection{Fast mPOD algorithms and computational cost}
\label{sec3b}

The compact-support structure described above enables the construction of the mPOD through reduced eigenvalue problems. In practice, this requires assembling the bandwise spectral correlation matrices $\mathbf K_{\mathcal F,m}$, which can be achieved through two alternative implementation routes.

In the first route, one follows the method-of-snapshots philosophy and works from the temporal correlation matrix $\mathbf K=\mathbf D^\dag \mathbf D$. The matrix $\mathbf K$ is transformed into the spectral domain to obtain $\mathbf K_{\mathcal F}$, after which the mask $\mathbf M_m$ is applied to extract the contribution of the $m$-th band. This route is natural when $n_t\ll n_s$, since the correlation matrix is then much smaller than the data matrix.

In the second route, one first transforms the data matrix into the spectral domain, obtaining $\widehat{\mathbf D}=\mathbf D\,\overline{\mathbf\Psi}_{\mathrm F}$, and then applies the mask directly to the Fourier coefficients of the data. The filtered spectral correlation matrix is then assembled as
\(\mathbf K_{\mathcal F,m}
=
\widehat{\mathbf D}_m^\dagger \widehat{\mathbf D}_m .
\)
This route avoids the construction of the full matrix $\mathbf K_{\mathcal F}$ and becomes advantageous when the spectral bands are narrow, i.e.\ when $n^{(m)}\ll n_t$, since all operations are restricted to the reduced spectral subspaces associated with the active frequencies.

The two implementations are equivalent in the ideal compact-support setting,
because the spectral masks enforce strictly disjoint frequency supports,
so that no cross-band interactions are present and the filtering operation
commutes with the construction of the correlation matrix. The overall procedure is summarized in Algorithm~\ref{alg:fastmpod}, while the two implementations of $\texttt{ComputeBandMatrix}$ are detailed in Algorithm~\ref{alg:bandmatrix}.

Importantly, the eigenvalue problem is not solved on the full matrix
$\mathbf K_{\mathcal F,m}\in\mathbb{C}^{n_t\times n_t}$, but only on its
compact nonzero blocks $\mathbf K_{\mathcal F,c}\in\mathbb{C}^{n^{(m)}\times n^{(m)}}$ introduced in the previous subsection.

This reduction is justified by the block-structured results in Appendix~\ref{sec:appendix}, which show that the nonzero eigenvalues and eigenvectors of $\mathbf K_{\mathcal F,m}$ are entirely determined by those of its compact block. Therefore, each scale leads to an independent eigenvalue problem whose size is determined solely by the number of active frequencies $n^{(m)}$, rather than the full temporal dimension $n_t$.

\begin{algorithm}[htb]
\caption{Fast spectral mPOD}
\label{alg:fastmpod}
\begin{algorithmic}[1]

\State \textbf{Input:} data matrix $\mathbf D$, frequency splitting vector $F_V$

\State Construct the spectral masks $\{M_m(f)\}_{m=1}^{n_M}$ from $F_V$.

\For{$m=1,\dots,n_M$}
\State Compute 
\(
\mathbf K_{\mathcal F,m}
=
\texttt{ComputeBandMatrix}(\mathbf D, \mathbf{M}_m)
\)
\State Extract compact block
\(
\mathbf K_{\mathcal F,c}
\)
\State Solve reduced eigenvalue problem
\(
\mathbf K_{\mathcal F,c}\,
\widehat{\mathbf \Psi}_m
=
\widehat{\mathbf \Psi}_m
\mathbf \Sigma_m^2
\)
\State Recover temporal modes
\(
\mathbf \Psi_m
=
\overline{\mathbf \Psi}_{\mathrm F}\widehat{\mathbf \Psi}_m
\)
\EndFor
\State Assemble temporal basis
\(
\mathbf \Psi_{\mathcal M}
=
[\mathbf \Psi_1,\dots,\mathbf \Psi_{n_M}]
\)

\State Compute spatial modes
\(
\mathbf \Phi_{\mathcal M}
=
\mathbf D\,\mathbf \Psi_{\mathcal M}\,
\mathbf \Sigma_{\mathcal M}^{-1}
\)

\end{algorithmic}
\end{algorithm}

\begin{algorithm}[htb]
\caption{\texttt{ComputeBandMatrix}$(\mathbf D,\mathbf{M}_m)$}
\label{alg:bandmatrix}
\begin{algorithmic}[1]


\State \textbf{Input:} data matrix $\mathbf D$, mask $\mathbf{M}_m$ of scale $m$


\If{$n_t \ll n_s$}
    \State Compute temporal correlation matrix
    \(
    \mathbf K=\mathbf D^\dag \mathbf D
    \)
    \State Compute spectral representation
    \(
    \mathbf K_{\mathcal F}
    =
    \mathbf \Psi_{\mathrm F}\,
    \mathbf K\,
    \overline{\mathbf \Psi}_{\mathrm F}
    \)
    \State Return
    \(
    \mathbf K_{\mathcal F,m}
    =
    \mathbf M_m\,
    \mathbf K_{\mathcal F}\,
    \mathbf M_m
    \)
\Else
    \State Compute Fourier-transformed data
    \(
    \widehat{\mathbf D}
    =
    \mathbf D\,\overline{\mathbf \Psi}_{\mathrm F}
    \)
    \State Extract filtered spectral data block
    \(
    \widehat{\mathbf D}_m
    =
    \widehat{\mathbf D}\,\mathbf M_m
    \)
    \State Return
    \(
    \mathbf K_{\mathcal F,m}
    =
    \widehat{\mathbf D}_m^\dagger \widehat{\mathbf D}_m
    \)
\EndIf

\end{algorithmic}
\end{algorithm}

The computational cost of the fast mPOD formulation is dominated by the construction of bandwise spectral matrices and by the reduced eigendecompositions. In all cases, extracting compact blocks from $\mathbf K_{\mathcal F,m}$ is negligible compared to the matrix products and eigensolves. 

For the correlation-based route (lines 4--6 in Algorithm~\ref{alg:bandmatrix}), the dominant steps are the construction of the temporal correlation matrix, with cost $O(n_s n_t^2)$, the spectral transformation of $\mathbf K$, with cost $O(n_t^2\log n_t)$, and the reduced eigendecompositions, with total cost \( O\!\left(\sum_{m=1}^{n_M} (n^{(m)})^3\right). \) The total complexity is therefore

\begin{equation} \label{eq:Kcompl} O\!\left( n_s n_t^2 + n_t^2\log n_t + \sum_{m=1}^{n_M}(n^{(m)})^3 \right). \end{equation} 

For the data-based route (lines 8--10 in Algorithm~\ref{alg:bandmatrix}), the Fourier transform of the data matrix costs $O(n_s n_t\log n_t)$, while the construction of the bandwise matrices \( \mathbf K_{\mathcal F,m}\), which costs \( O\!\left(\sum_{m=1}^{n_M} n_s (n^{(m)})^2\right). \) Adding the reduced eigendecompositions gives the total complexity 

\begin{equation} \label{eq:Dcompl} O\!\left( n_s n_t\log n_t + \sum_{m=1}^{n_M} n_s (n^{(m)})^2 + \sum_{m=1}^{n_M}(n^{(m)})^3 \right). \end{equation}

A similar acceleration can be applied to the spatial-correlation formulation based on $\mathbf C_m=\mathbf D_m \mathbf D_m^\dagger$. In this case, the compact-support structure allows one to compute
\(\mathbf C_m = \widehat{\mathbf D}_m \widehat{\mathbf D}_m^\dagger\) so that only the $n^{(m)}$ active spectral coefficients are involved. This reduces the cost of assembling $\mathbf C_m$ from $O(n_s^2 n_t)$ to $O(n_s^2 n^{(m)})$ per band. However, the size of the associated eigenvalue problems remains $n_s\times n_s$, so that the overall computational gain is less pronounced than in the temporal-correlation formulation.
\section{Verification and Performance Assessment}\label{sec4}

This section verifies that the proposed formulation reproduces the results of the classical mPOD while significantly reducing computational cost. Section~\ref{sec4p1} compares the scale-separation obtained with the classical FIR-based filters and the proposed spectral-mask formulation using a synthetic test case. Section~\ref{sec4p2} demonstrates the performance of the method on a canonical fluid-dynamics dataset consisting of the wake of a circular cylinder at Reynolds number $\mbox{Re} \approx 5000$. Finally, Section~\ref{sec4p3} analyzes the computational scaling of the proposed algorithm and compares it with the classical mPOD implementation.

\subsection{Synthetic test: FIR filters vs spectral masks}\label{sec4p1}

This test compares the mask-based fast mPOD approach to the classical FIR filter-based approach on an artificial dataset. The test has three main objectives: (i) to compare the shapes of the individual masks $M_m(f)$ with the amplitudes of the FIR filter responses $| \widehat{\mathbf H}_m(f) |$, (ii) to evaluate the reconstruction properties of the masks and filters, i.e.\ $\sum_{m=1}^{n_M} M_m(f)$ and $\sum_{m=1}^{n_M} \widehat{\mathbf H}_m(f)$, and (iii) to assess their performance in mitigating the Gibbs phenomenon. For this last objective, three approaches are compared: (i) the FIR filter-based formulation as in Algorithm~\ref{alg:classicK}, (ii) the mask-based approach as in Algorithm~\ref{alg:fastmpod}, and (iii) a variant of the mask-based approach with infinitely sharp band boundaries.

\begin{figure*}[htbp]
    \centering
    \includegraphics[width = 0.75\textwidth]{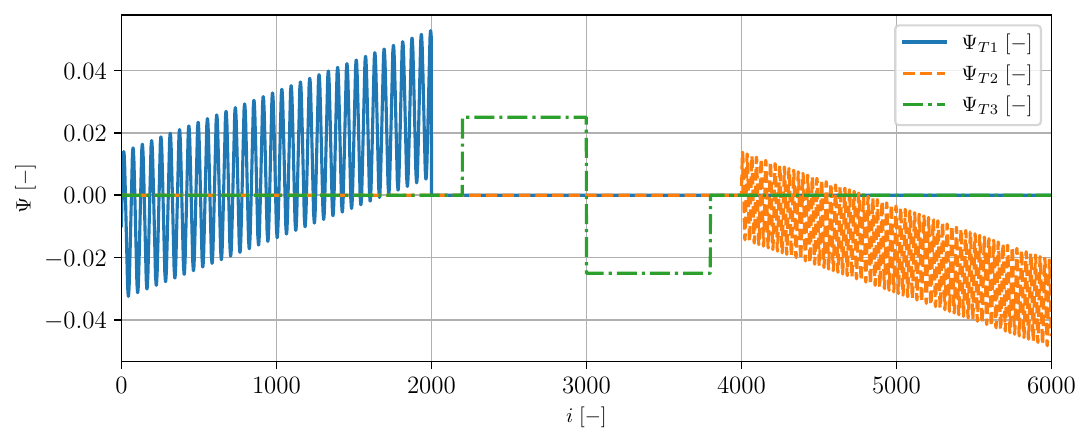}
    \caption{Temporal functions used to assemble the test dataset. Test mode 1 ($\Psi_{T1}$) is nonzero only in the range temporal indices $i \in (0, 2000)$ with the slope defined as $(i-500)/1200$ and the superimposed sine wave with $f_1/f_s = 1/60$. Test mode 2 ($\Psi_{T2}$) is nonzero only in the range temporal indices $i \in (4000, 6000)$ with the slope defined as $(4000-i)/800$ and the superimposed sine wave with $f_2/f_s = 1/30$. The third test mode ($\Psi_{T3}$) is a single wavelet in the interval $i \in (2200,3800)$.}
    \label{fig:chronoses}
\end{figure*}

The test is based on the mPOD analysis of an artificial dataset with $n_t = 6000$, which was constructed to pose a challenging problem for Fourier-based approaches, thereby promoting the appearance of the Gibbs phenomenon. The temporal evolution shapes used to create the test data are shown in Figure~\ref{fig:chronoses}, with the sine waves superimposed on the slopes having frequencies of $f_1/f_s = 1/60$ and $f_2/f_s = 1/30$, respectively.   
The mPOD analysis was performed using the frequency splitting vector $F_V/f_s = [1/120,3/120,5/120,10/120]$ with the taper region size set to $\delta/f_s = 1/300$ for all scales and the FIR filter order set to 1001 for all scales. 

\begin{figure}[htbp]
    \centering
    \includegraphics[width = 0.48\textwidth]{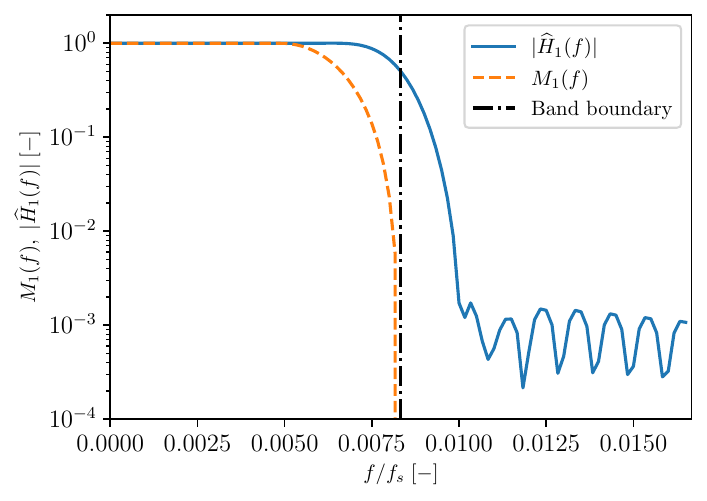}
    \caption{Mask and the magnitude of the low-pass FIR filter frequency response of the first frequency band.}
    \label{fig:band1}
\end{figure}

\begin{figure*}[hbtp]
    \centering
    \includegraphics[width = 0.75\textwidth]{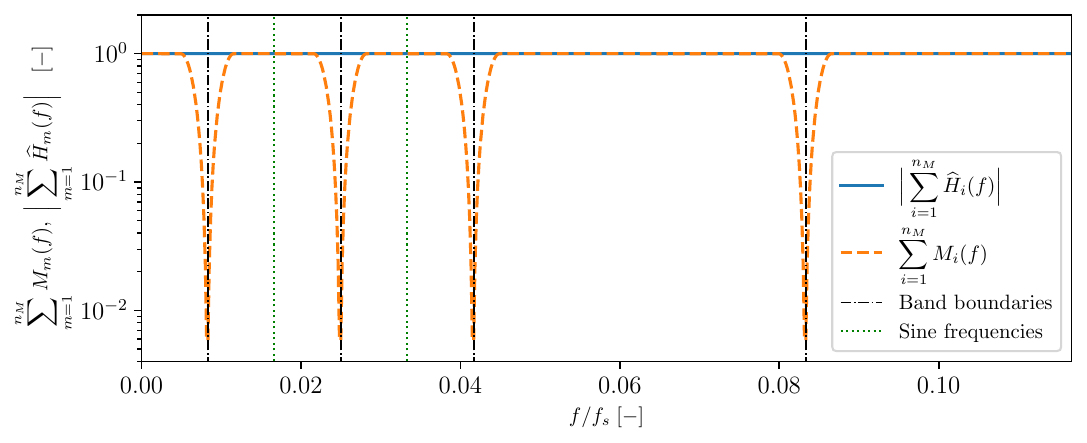}
    \caption{Reconstruction property of the masks and FIR filters. The drops in the mask-based approach around the frequency block boundaries due to the combination of smooth taper and disjoint spectral supports of the masks are clearly visible.}
    \label{fig:recProp}
\end{figure*}

As shown in Figure~\ref{fig:band1}, the mask exhibits a taper shape similar to that of the FIR filter amplitude. However, unlike the FIR filters, the taper of the mask is entirely confined within the corresponding frequency band, as described in Section~\ref{sec3}. This ensures perfect spectral separation, but at the expense of the reconstruction property, as illustrated in Figure~\ref{fig:recProp}.

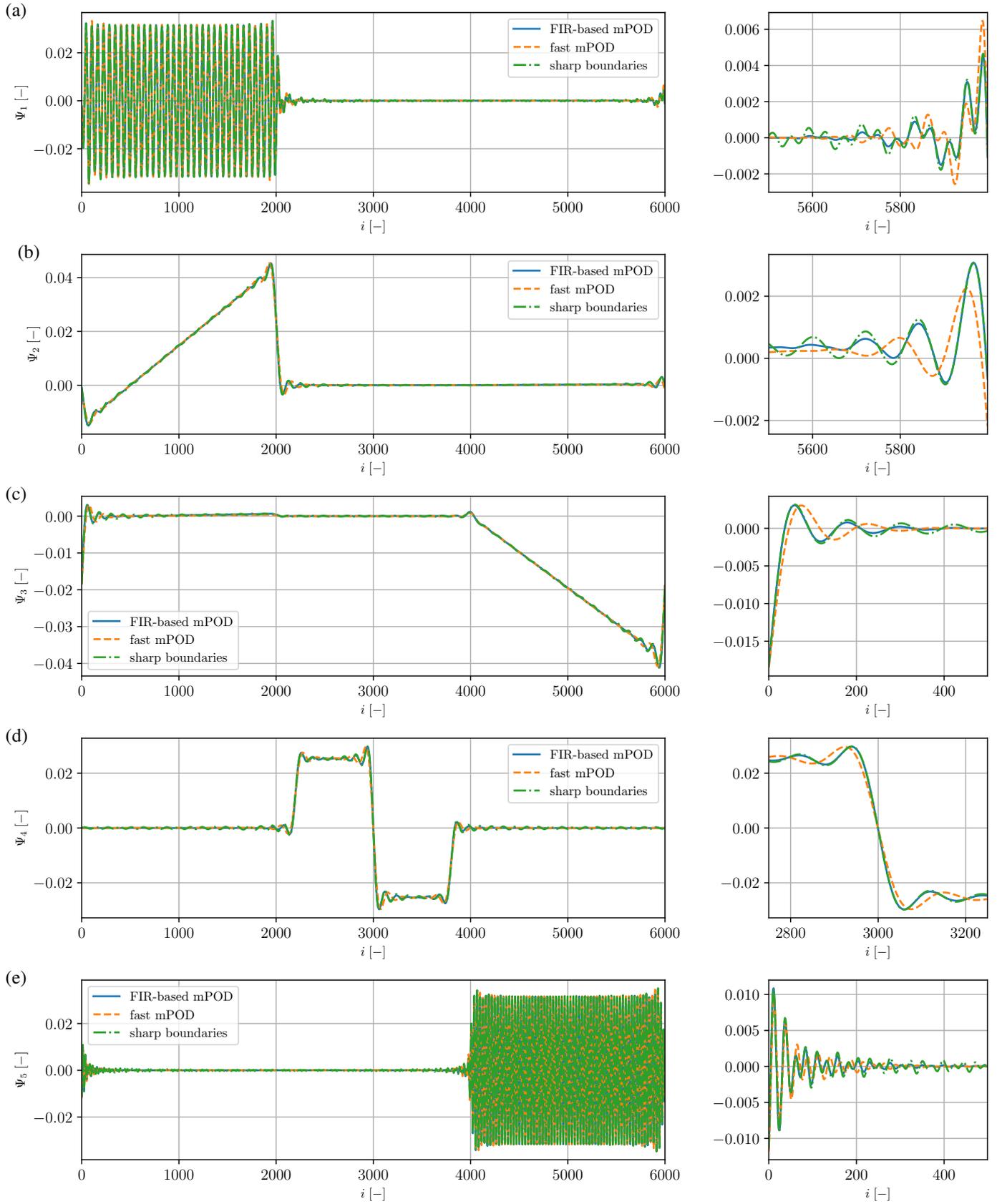
\begin{figure*}[htbp]
    \centering
    \begin{tikzpicture}
        \node[anchor=north west] (subfig1) at (0,0) {\includegraphics[height=0.185\textheight]{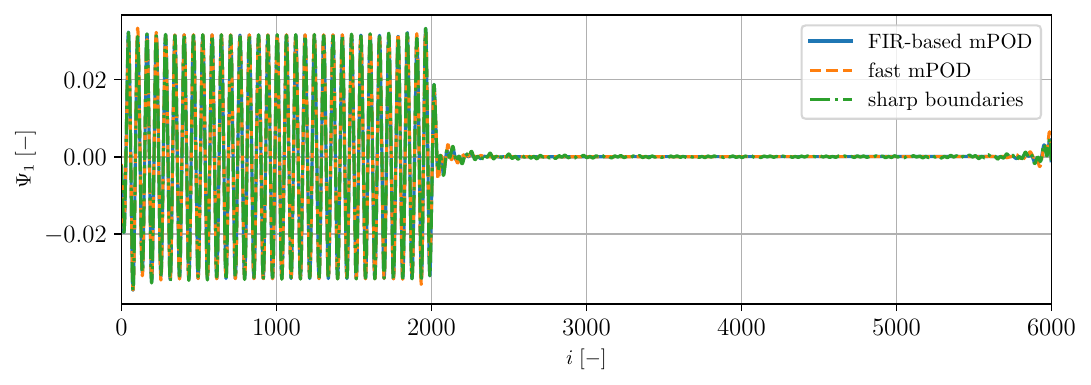}};
        \node[anchor=north west] at (subfig1.north west) {(a)};

        \node[anchor=north east] (subfig2) at ([yshift=0.2cm]subfig1.south east) {\includegraphics[height=0.185\textheight]{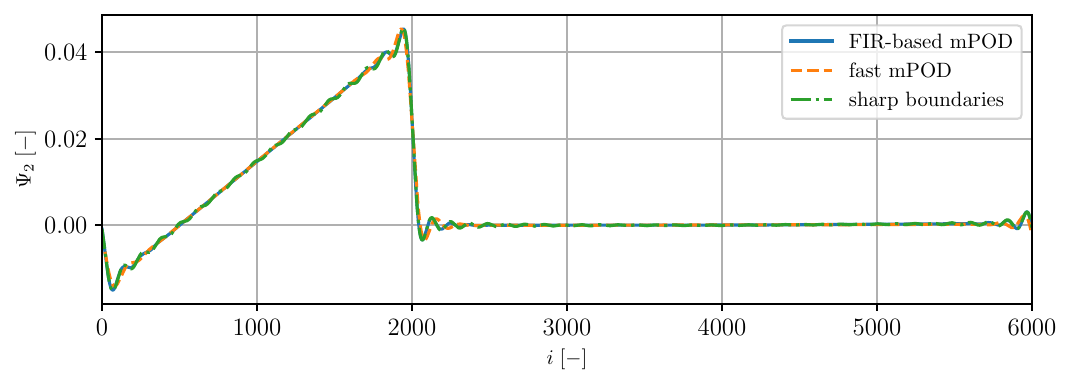}};
        \node[anchor=north west] at (subfig2.north west) {(b)};

        \node[anchor=north east] (subfig3) at ([yshift=0.2cm]subfig2.south east) {\includegraphics[height=0.185\textheight]{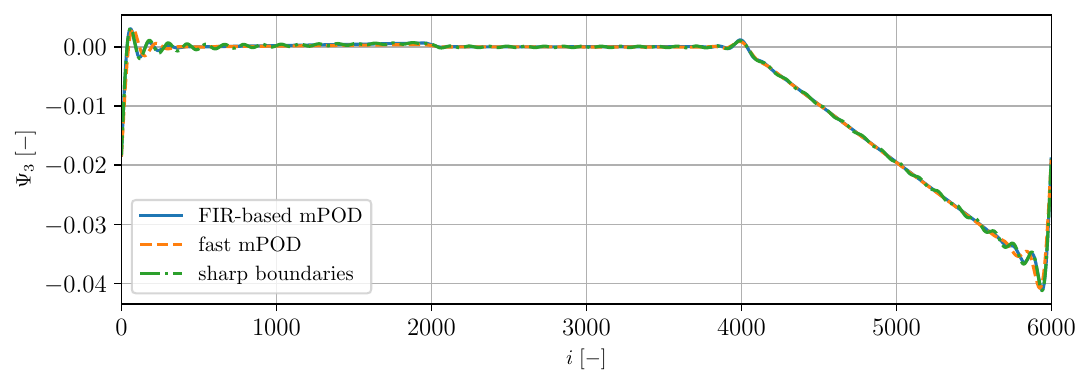}};
        \node[anchor=north west] at (subfig3.north west) {(c)};

        \node[anchor=north east] (subfig4) at ([yshift=0.2cm]subfig3.south east) {\includegraphics[height=0.185\textheight]{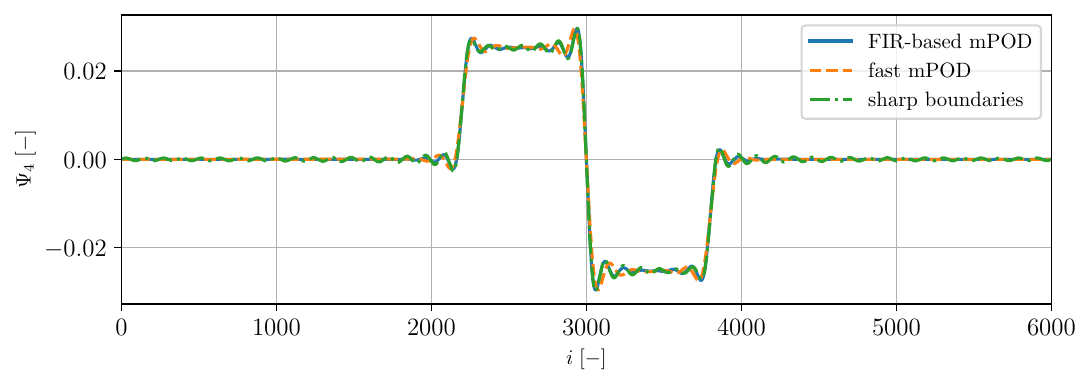}};
        \node[anchor=north west] at (subfig4.north west) {(d)};

        \node[anchor=north east] (subfig5) at ([yshift=0.2cm]subfig4.south east) {\includegraphics[height=0.185\textheight]{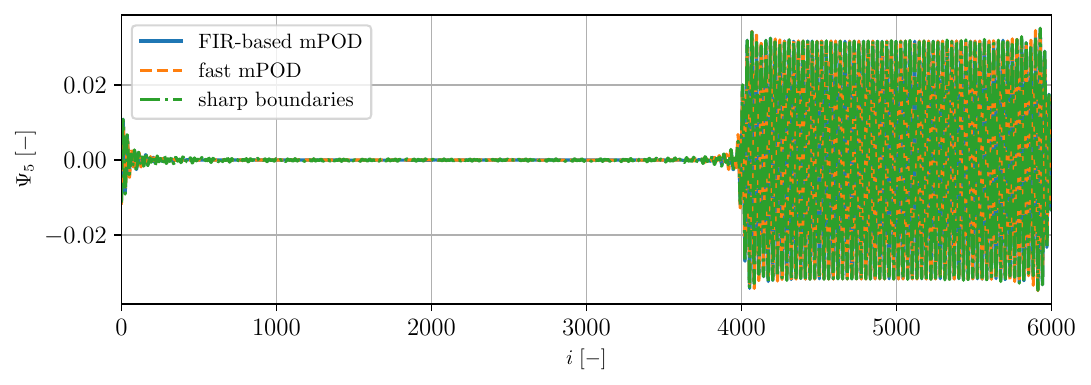}};
        \node[anchor=north west] at (subfig5.north west) {(e)};

        \node[anchor=north west] (subfig1det) at (subfig1.north east){\includegraphics[height=0.185\textheight]{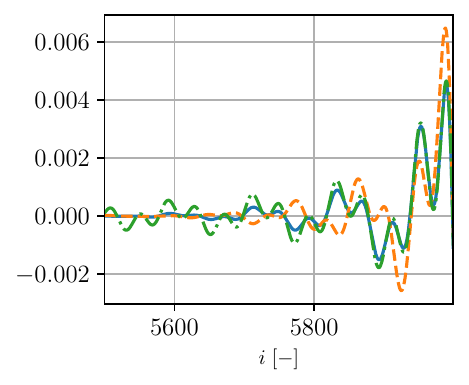}};
        \node[anchor=north east] (subfig2det) at ([yshift=0.2cm]subfig1det.south east){\includegraphics[height=0.185\textheight]{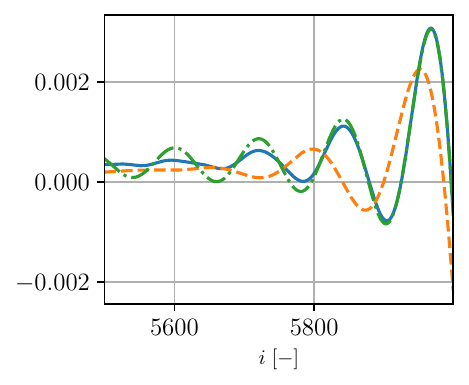}};
        \node[anchor=north east] (subfig3det) at ([yshift=0.2cm]subfig2det.south east){\includegraphics[height=0.185\textheight]{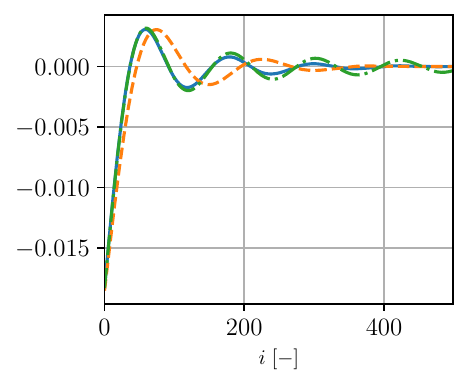}};
        \node[anchor=north east] (subfig4det) at ([yshift=0.2cm]subfig3det.south east){\includegraphics[height=0.185\textheight]{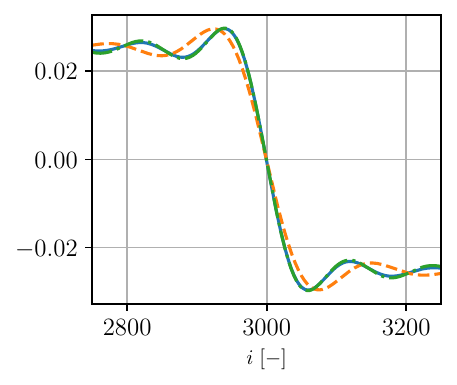}};
        \node[anchor=north east] (subfig5det) at ([yshift=0.2cm]subfig4det.south east){\includegraphics[height=0.185\textheight]{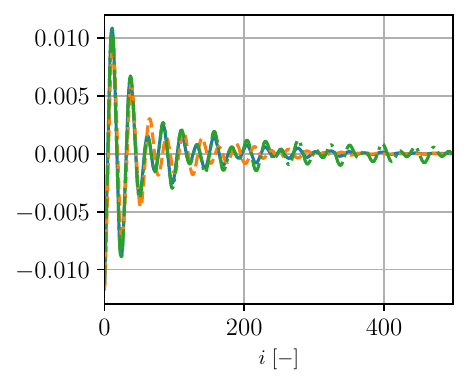}};

    \end{tikzpicture}
    \caption{mPOD modes from the decomposition of the artificial dataset to showcase the Gibbs phenomenon. All 5 significant modes together with the relevant details are shown. The figure is structured as follows: (a) Mode 1, (b) Mode 2, (c) Mode 3, (d) Mode 4, and (e) Mode 5. The frequency split of the mPOD allowed for isolation of the sine waves while retaining the temporal localization of those waves.}
    \label{fig:test1Res}
\end{figure*}

The impact of this trade-off on the Gibbs phenomenon is illustrated in Figure~\ref{fig:test1Res}. The mask-based approach exhibits localized oscillations near temporal discontinuities, which are slightly more pronounced than in the FIR-based formulation due to the absence of spectral overlap. However, these oscillations remain significantly smaller than those obtained with sharp spectral truncation, demonstrating that the smooth tapering within each band effectively mitigates spurious ringing while preserving strict spectral separation.

 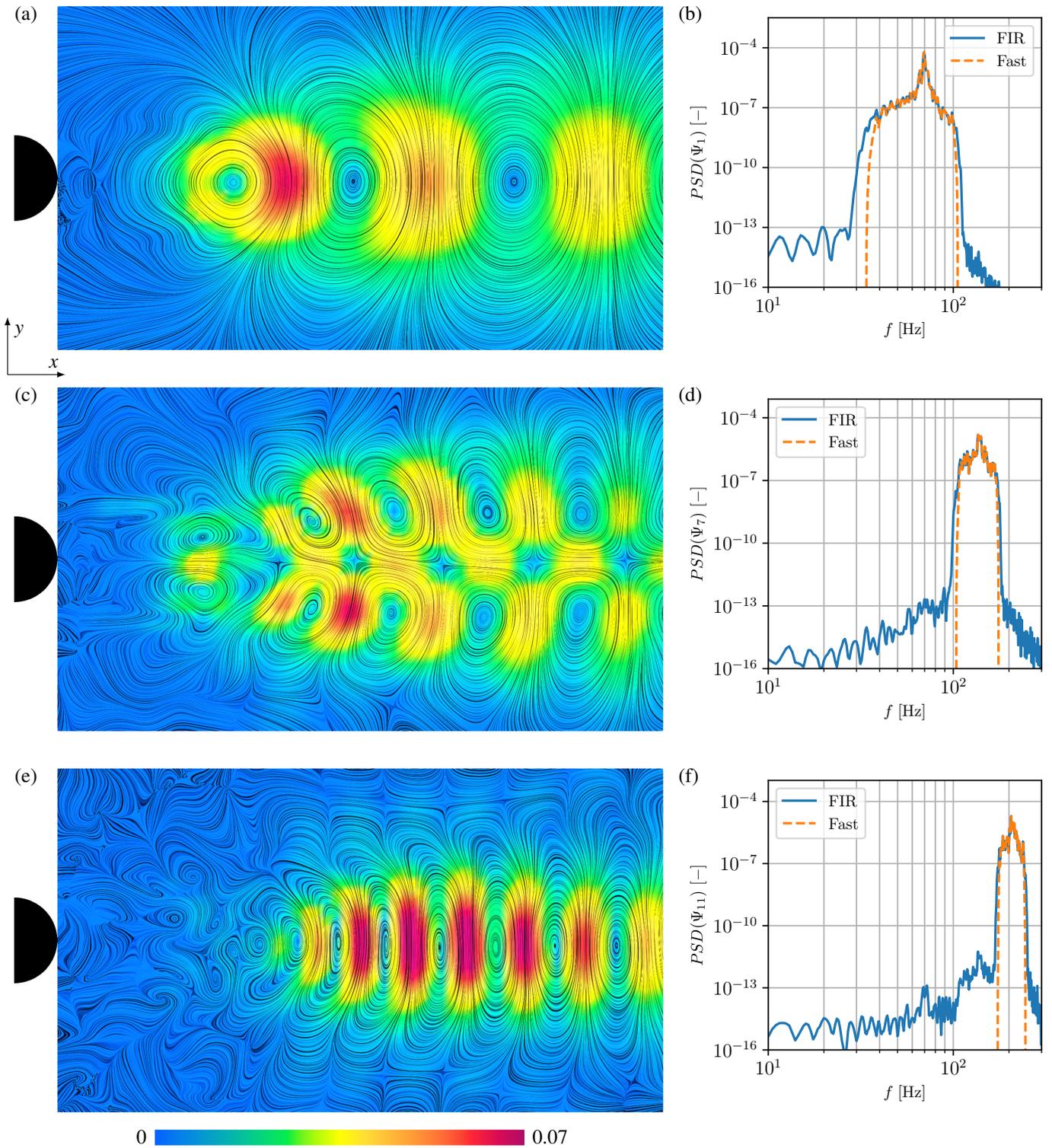
\begin{figure*}[htbp]
    \centering
    \begin{tikzpicture}
        \node[anchor=north west] (subfig1) at (0,0) {\includegraphics[height=0.25\textheight]{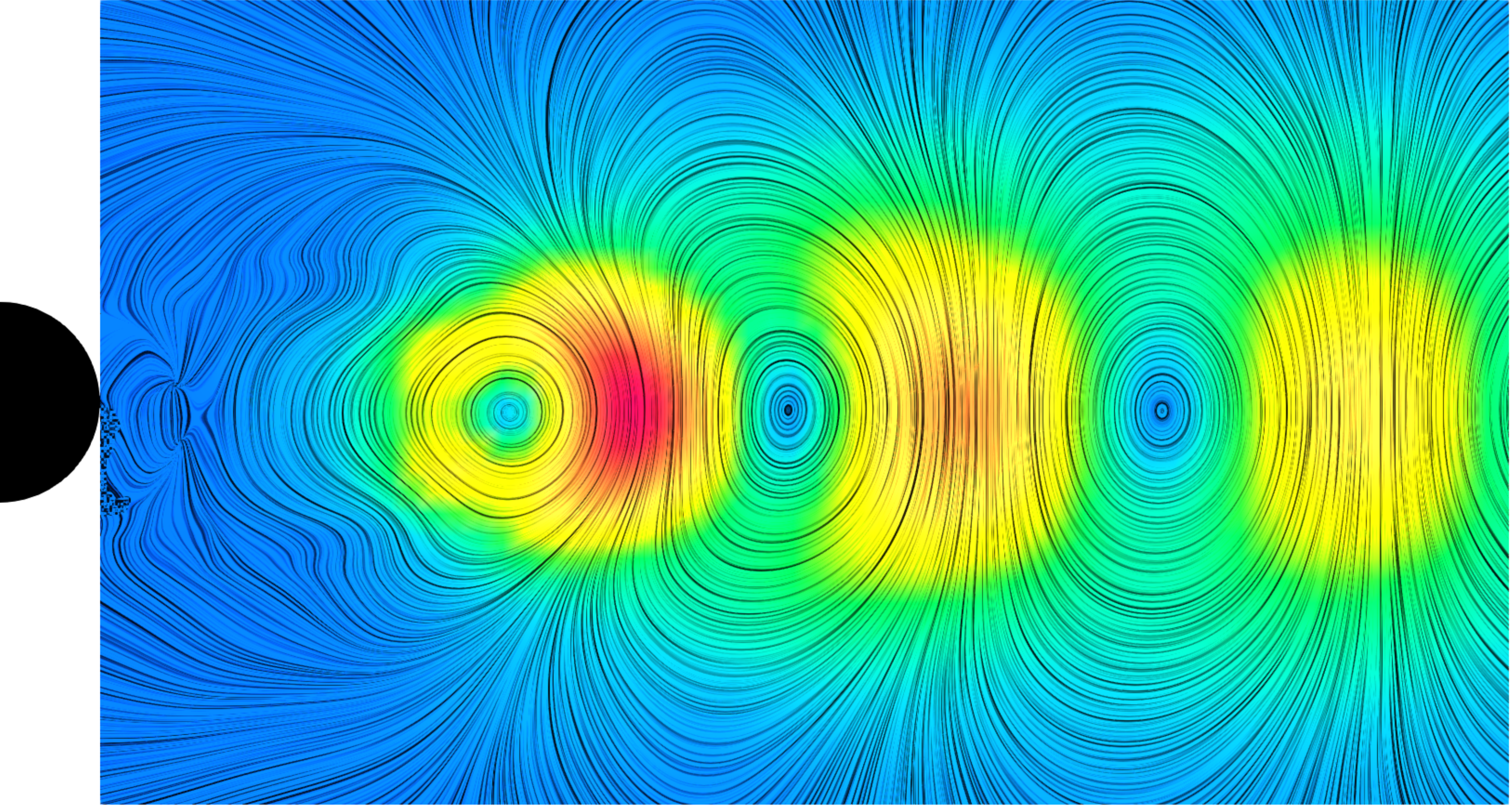}};
        \node[anchor=north west] at (subfig1.north west) {(a)};
        \draw[-latex] ([yshift=0-0.3cm]subfig1.south west) -- ([yshift=0.7cm]subfig1.south west) node[pos=1.0,anchor=north west] {$y$};
        \draw[-latex] ([yshift=0-0.3cm]subfig1.south west) -- ([xshift=1cm, yshift=0-0.3cm]subfig1.south west) node[pos=1.0,anchor=south east] {$x$};

        \node[anchor=north west] (subfig2) at (subfig1.north east) {\includegraphics[height=0.25\textheight]{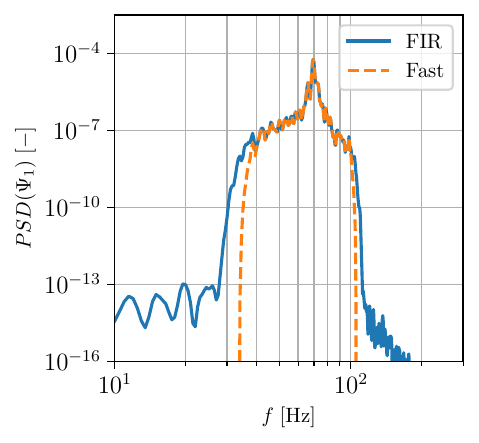}};
        \node[anchor=north west] at (subfig2.north west) {(b)};

        \node[anchor=north east] (subfig3) at ([yshift=0-0.4cm]subfig1.south east) {\includegraphics[height=0.25\textheight]{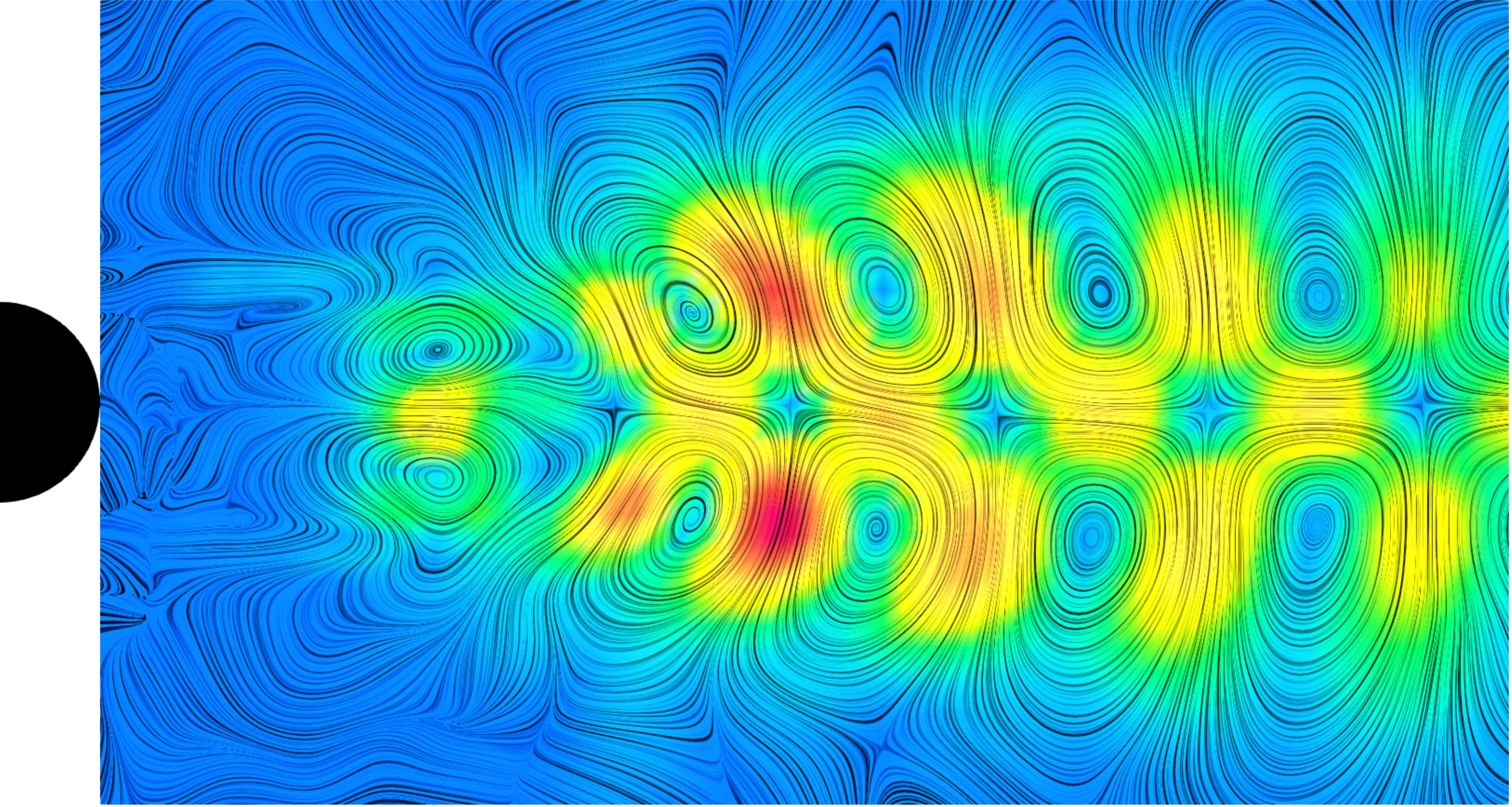}};
        \node[anchor=north west] at (subfig3.north west) {(c)};

        \node[anchor=north west] (subfig4) at (subfig3.north east) {\includegraphics[height=0.25\textheight]{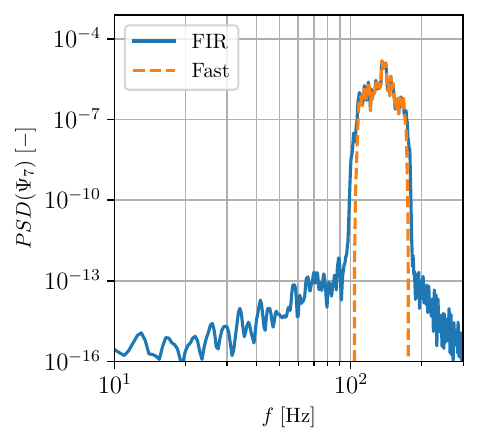}};
        \node[anchor=north west] at (subfig4.north west) {(d)};

        \node[anchor=north east] (subfig5) at ([yshift=0-0.4cm]subfig3.south east) {\includegraphics[height=0.25\textheight]{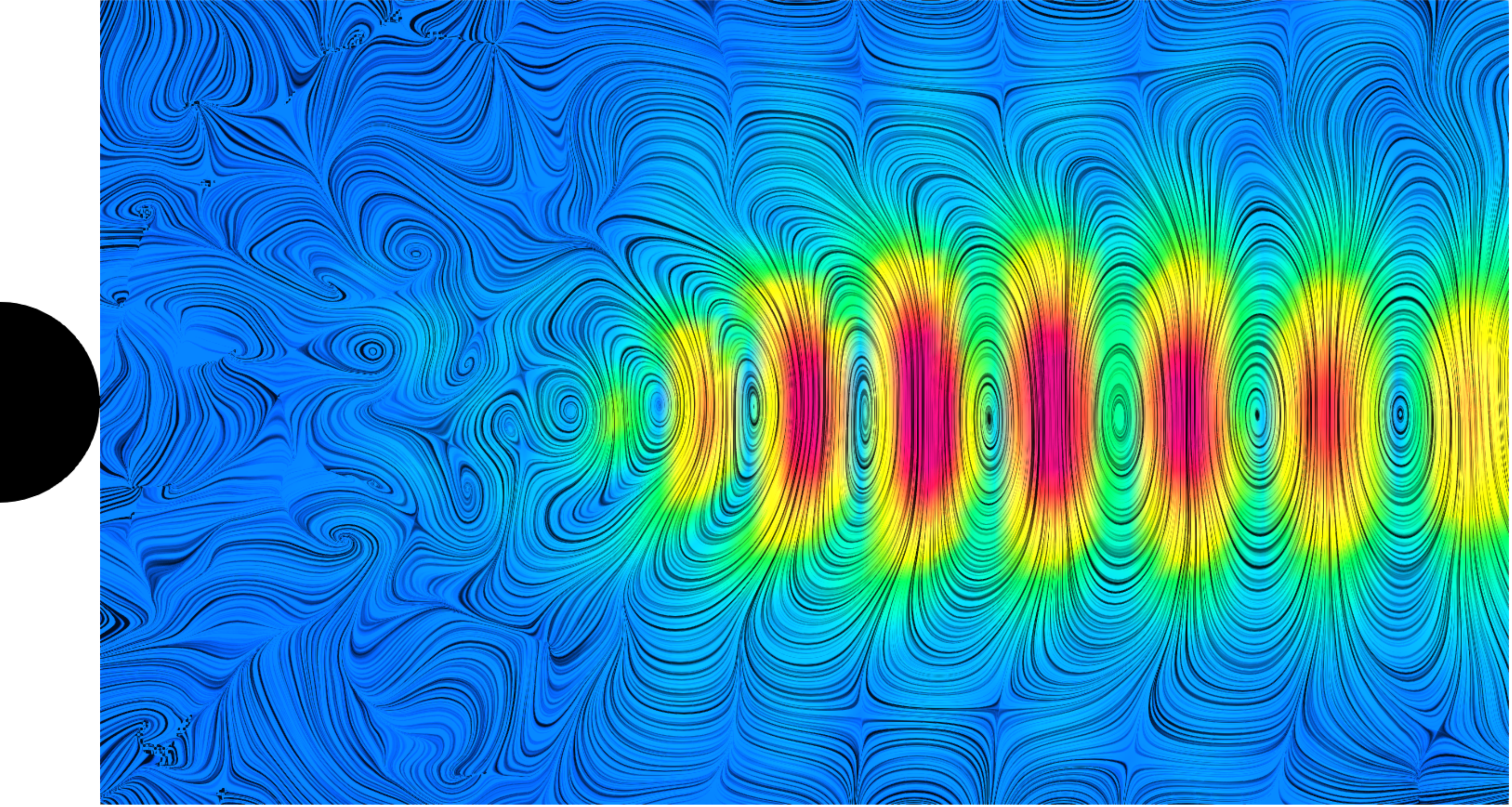}};
        \node[anchor=north west] at (subfig5.north west) {(e)};

        \node[anchor=north west] (subfig6) at (subfig5.north east) {\includegraphics[height=0.25\textheight]{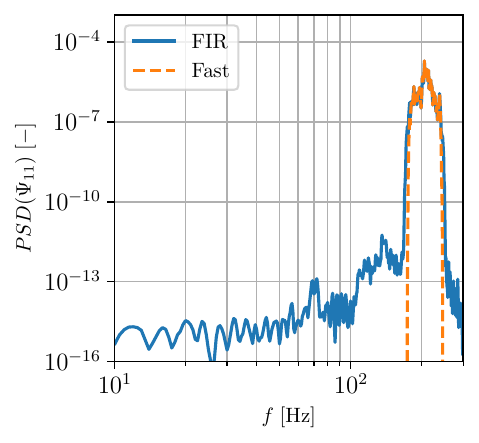}};
        \node[anchor=north west] at (subfig6.north west) {(f)};

        \node[anchor=north] (legMode) at (subfig5.south) {\includegraphics[height=10px, width=0.35\textwidth]{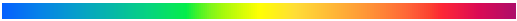}};
        \node[anchor=east,outer sep=0pt,inner sep=0pt] at (legMode.west) {0};
        \node[anchor=west,outer sep=0pt,inner sep=0pt] at (legMode.east) {$0.07$};

    \end{tikzpicture}
    \caption{mPOD modes corresponding to first (a, b), second (c, d), and third (e, f) harmonic frequency, (a, c, e) magnitude with line integral convolution (LIC) streamlines for visualizing the flow patterns, (b, d, f) corresponding mode spectra}
    \label{fig:modes}
\end{figure*}

 \begin{figure*}[htbp]
    \centering
    \begin{tikzpicture}
        \node[anchor=north west] (subfig1) at (0,0) {\includegraphics[height=0.28\textheight]{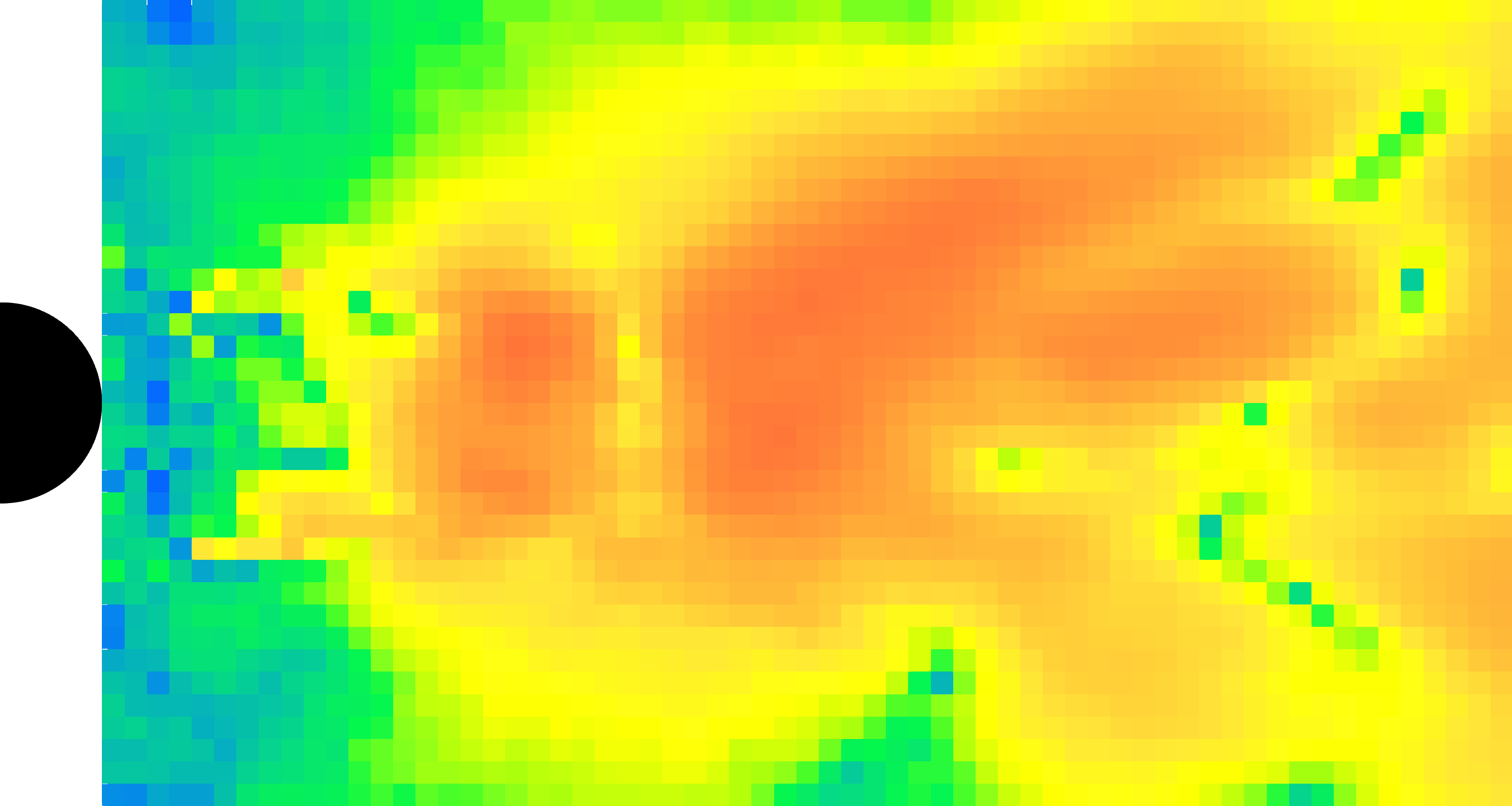}};
        \node[anchor=north west] at (subfig1.north west) {(a)};
        \draw[-latex] ([yshift=0-0.3cm]subfig1.south west) -- ([yshift=0.7cm]subfig1.south west) node[pos=1.0,anchor=north west] {$y$};
        \draw[-latex] ([yshift=0-0.3cm]subfig1.south west) -- ([xshift=1cm, yshift=0-0.3cm]subfig1.south west) node[pos=1.0,anchor=south east] {$x$};


        \node[anchor=north] (subfig3) at ([yshift=0-0.4cm]subfig1.south) {\includegraphics[height=0.28\textheight]{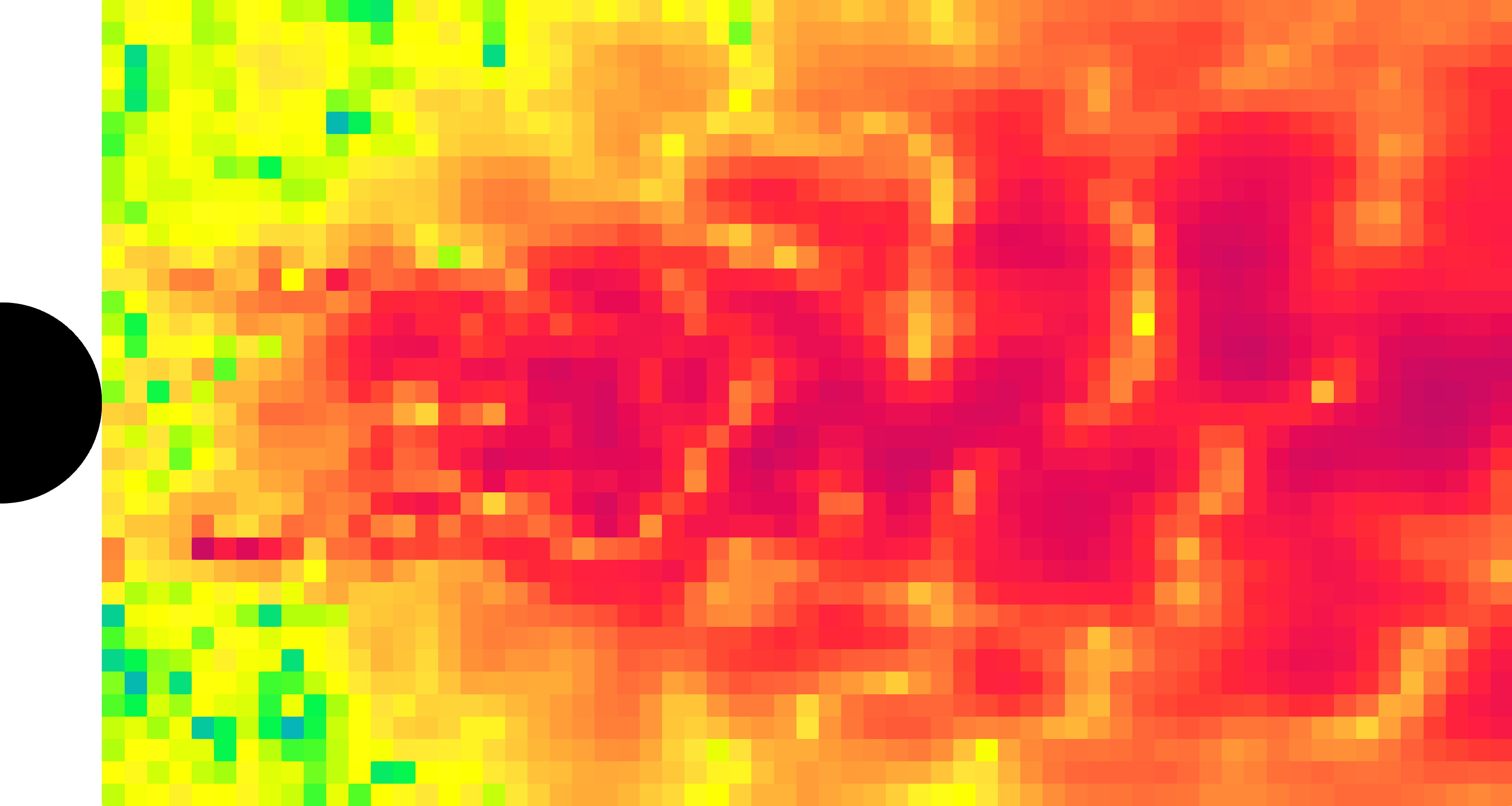}};
        \node[anchor=north west] at (subfig3.north west) {(b)};


        \node[anchor=north] (subfig5) at ([yshift=0-0.4cm]subfig3.south) {\includegraphics[height=0.28\textheight]{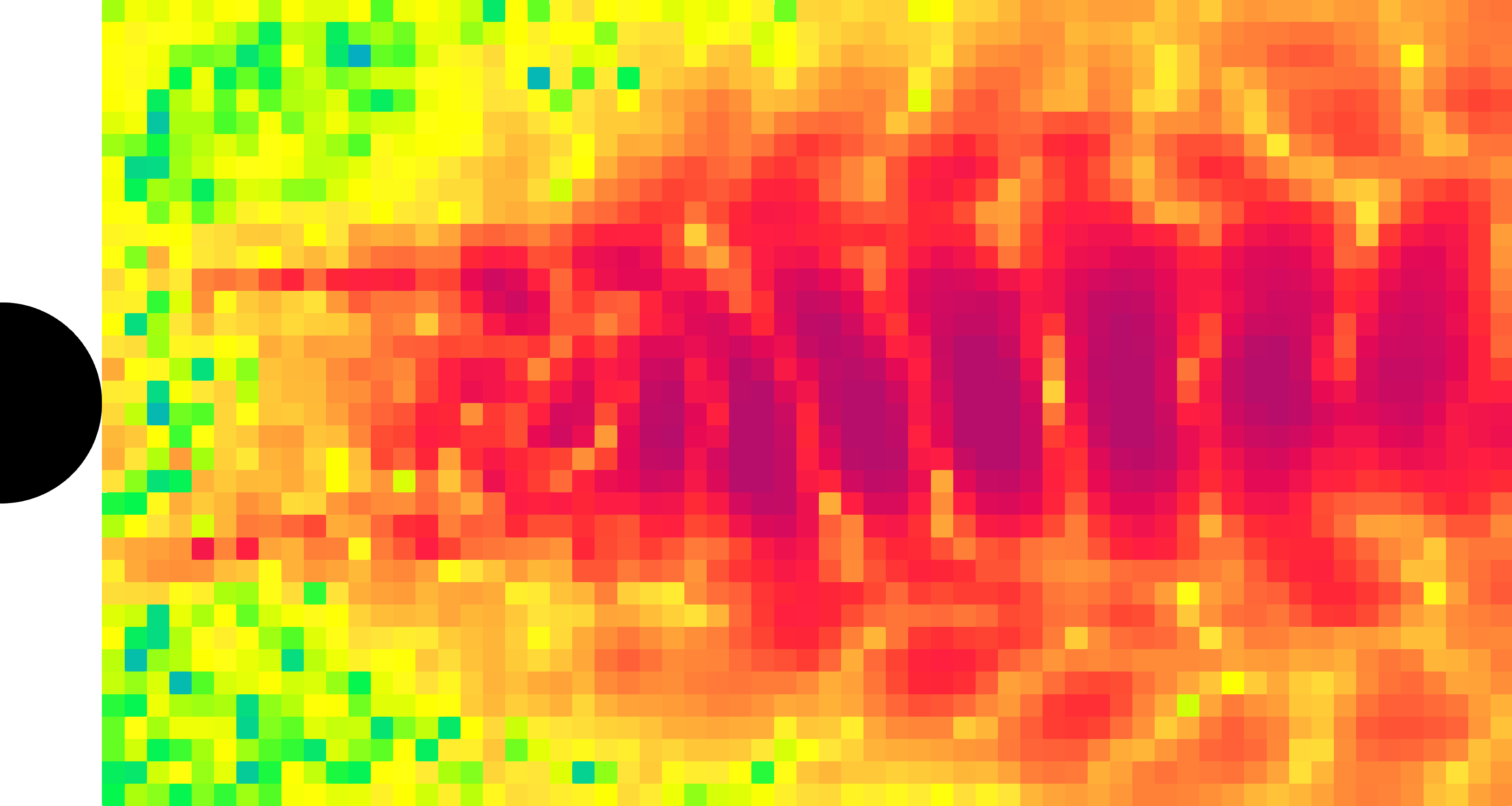}};
        \node[anchor=north west] at (subfig5.north west) {(c)};


        \node[anchor=north] (legMode3L) at (subfig5.south) {\includegraphics[height=10px, width=0.33\textwidth]{rainbowUniHor.png}};
        \node[anchor=east,outer sep=0pt,inner sep=0pt] at (legMode3L.west) {$10^{-7}$};
        \node[anchor=west,outer sep=0pt,inner sep=0pt] at (legMode3L.east) {$10^{-3}$};

        
    \end{tikzpicture}
    \caption{Absolute differences between FIR implementation and fast mPOD modes. Modes correspond to (a) first, (b) second, (c) third harmonic frequency.}
    \label{fig:diff}
\end{figure*}

\begin{figure*}[t]
    \centering
    \begin{tikzpicture}
        \node[anchor=north west] (subfig1) at (0,0) {\includegraphics[width=0.7\textwidth]{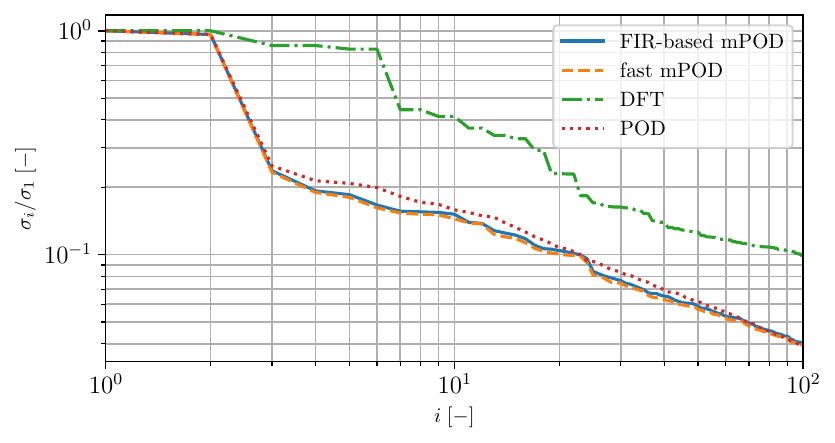}};

    \end{tikzpicture}
    \caption{Convergence of different data decomposition methods on the cylinder-in-crossflow dataset.}
    \label{fig:conv}
\end{figure*}


\subsection{A classic experimental benchmark}\label{sec4p2}

The experimental dataset used in this study is taken from~\citep{Hlavaty2024}, where the canonical case of a cylinder in crossflow is investigated using time-resolved particle image velocimetry (TR-PIV). This configuration, being one of the standard benchmarks in fluid dynamics, is adopted here to assess the performance of the proposed method. For a complete description of the experimental setup and measurement procedure, the reader is referred to~\citep{Hlavaty2024}.

Measurements were conducted in a wind tunnel with a closed test section of a $250 \times 250 \, \mathrm{mm^2}$ cross-section. The free-stream velocity was $U_{in} = 5 \, \mathrm{m/s}$ and the cylinder diameter was $d = 15 \, \mathrm{mm}$, yielding a Reynolds number $
\mbox{Re} \approx 5000$ and a fundamental vortex shedding frequency of $f_{1H} \approx 70 \, \mathrm{Hz}$. The experiments were performed in air at ambient conditions ($\nu = 1.5 \cdot 10^{-5} \, \mathrm{m^2/s}$, $\rho = 1.2 \, \mathrm{kg/m^3}$), with the cylinder oriented transversely to the flow and spanning the entire test section.

The velocity fields were acquired in a plane normal to the cylinder axis using TR-PIV with a sampling frequency of $f_s = 2 \, \mathrm{kHz}$. The region of interest has a spatial resolution of $n_x = 63$ and $n_y = 36$ points in the $x$ and $y$ directions, respectively, with two velocity components per point, resulting in a total spatial dimension of $n_s = 4536$. A total of 2 seconds of data were recorded, corresponding to $n_t=4000$ snapshots.

The mPOD processing was based on the frequency splitting vector $F_V = [35, 105, 175, 245]$ Hz, with the last scale (above 245 Hz) being discarded. This places the band boundaries between successive harmonic frequencies. The FIR filter order for the classical mPOD was set to 501 for all scales, while the taper width for the fast mPOD was set to $\delta = f_{1H}/10 = 7 \, \mathrm{Hz}$. In all cases, the algorithm was configured to extract the 10 most energetic modes. Figure~\ref{fig:modes} presents the spatial structures and corresponding spectra of the modes associated with the first, second, and third harmonic frequencies.

Figure~\ref{fig:diff} shows the absolute differences between the classical and fast mPOD results, displayed on a logarithmic color scale. Overall, the fast mPOD accurately reproduces the spatial structures obtained with the classical formulation, with relative differences on the order of a few percent. These discrepancies are predominantly confined to regions of low fluctuation intensity, where the flow does not exhibit coherent structures and the modal amplitudes are small. 

In contrast, the dominant coherent structures associated with the vortex shedding are recovered almost identically. This indicates that the proposed formulation preserves the relevant content of the decomposition, while the observed differences mainly affect low-energy components. The remaining discrepancies can be attributed to two closely related factors: (i) the absence of spectral overlap between adjacent bands in the fast formulation, and (ii) the QR orthogonalization step in the classical mPOD, which slightly modifies the temporal modes to enforce orthogonality, with a corresponding propagation to the spatial modes.

Figure~\ref{fig:conv} compares the convergence behavior of the different decompositions in terms of singular value decay. The results show that mPOD retains the rapid convergence characteristics of POD while enforcing spectral separation across scales. In particular, the decay of the modal amplitudes closely follows that of POD, indicating that the energy-optimality of the decomposition is largely preserved within each band. By contrast, the DFT exhibits a much slower decay, as expected from its purely frequency-based nature, which does not provide an optimal energy ranking of the modes. This results highlights the ability of mPOD to combine spectral localization with efficient energy capture, even in flows with strongly periodic dynamics.

\subsection{Computational complexity and scaling}\label{sec4p3}

 \begin{figure*}[t]
    \centering
    \begin{tikzpicture}
        \node[anchor=north west] (subfig1) at (0,0) {\includegraphics[width=0.485\textwidth]{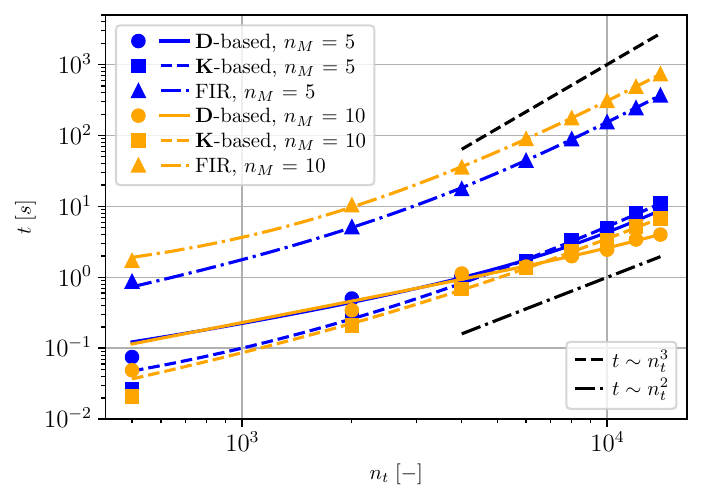}};
        \node[anchor=north west] at (subfig1.north west) {(a)};

        \node[anchor=north west] (subfig2) at (subfig1.north east) {\includegraphics[width=0.485\textwidth]{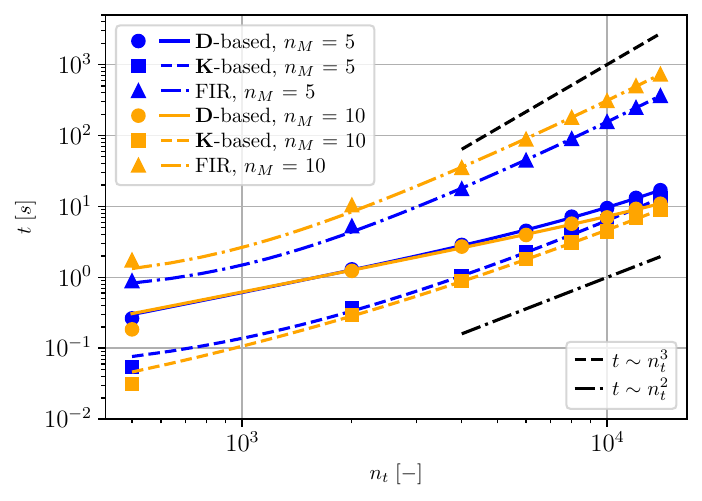}};
        \node[anchor=north west] at (subfig2.north west) {(b)};

    \end{tikzpicture}
    \caption{Computation times for (a) $n_s = 3000$ and (b) $n_s = 12000$. Approaches according to Algorithm~\ref{alg:classicK}, correlation-based version of Algorithm~\ref{alg:fastmpod}, and data-based version of Algorithm~\ref{alg:fastmpod} are compared. In both (a) and (b), it can be seen that the correlation-based approach is good for slender data matrices, while the data-based approach is best suited for short and fat data matrices.}
    \label{fig:complexity}
\end{figure*}

Computational complexity was assessed on synthetic test cases using random data matrices and a uniform frequency partition. The dependence of computation time on the number of temporal snapshots was analyzed for four configurations, obtained by combining two spatial dimensions, $n_s = 3000$ and $n_s = 12000$, with two numbers of frequency bands, $n_M = 5$ and $n_M = 10$. The bands were uniformly distributed over the frequency range, yielding $n^{(m)} \approx n_t/n_M$ for all $m$. For each configuration, the number of snapshots was varied in the range $n_t \in [5 \cdot 10^2, 2 \cdot 10^3, 4 \cdot 10^3, 6 \cdot 10^3, 8 \cdot 10^3, 10^4, 1.2 \cdot 10^4]$. The results shown in Figure~\ref{fig:complexity} correspond to averages over four realizations, computed using the MODULO toolbox (version 2.1.6) on Intel Xeon Gold 3.0 GHz processors using 12 cores.

As shown in Figure~\ref{fig:complexity}, the fast mPOD achieves speedups of up to two orders of magnitude compared to the classical formulation. Moreover, a distinctive scaling behavior of the fast mPOD is observed: the computational cost decreases as the number of frequency bands $n_M$ increases. This behavior follows directly from the reduced eigenvalue problems, whose dimensions scale as $n^{(m)} \approx n_t/n_M$. Substituting this relation into the complexity estimates in equations~\ref{eq:Kcompl} and~\ref{eq:Dcompl} yields contributions proportional to $1/n_M$ and $1/n_M^{2}$, respectively.

This trend contrasts with the classical mPOD, whose computational cost increases linearly with $n_M$, as shown in equation~\ref{eq:FIRCompl}. Therefore, while increasing the number of scales typically leads to higher costs in the classical formulation, it instead results in cheaper computations in the fast formulation due to the progressive reduction of the spectral subspaces.

Finally, comparing the two implementations of the fast mPOD shows that the correlation-based approach is most efficient for tall matrices ($n_t \ll n_s$), whereas the data-based approach performs best for short, wide matrices. This behavior is consistent with the theoretical complexity estimates and reflects the different matrix dimensions involved in each formulation.
\section{Conclusion}
\label{sec5}

This work revisits the multiscale Proper Orthogonal Decomposition (mPOD) with a focus on its computational complexity and the identification of its main bottlenecks. In the classical formulation based on FIR filters, the scale correlation matrices concentrate their energy within prescribed frequency bands but are not strictly low-rank, which necessitates solving full-order eigenvalue problems at each scale.

To overcome this limitation, we introduced a fast spectral formulation of the mPOD based on masks with compact spectral support. By enforcing strictly disjoint frequency bands, the proposed approach yields scale-specific correlation matrices that are exactly low-rank and admits a block structure that allows the eigenvalue problems to be solved in reduced spectral subspaces. This results in a significant reduction of the computational cost.

The method was assessed on both synthetic and experimental datasets. The synthetic test case highlighted the trade-off between spectral separation and Gibbs mitigation: while the fast mPOD exhibits slightly stronger oscillations than the FIR-based formulation, it remains far superior to sharp spectral truncation. The experimental test on a cylinder wake at Reynolds $\mbox{Re} \approx 5000$ demonstrated that the fast mPOD accurately reproduces the spatial structures and spectral content of the classical formulation, with only minor differences confined to low-energy regions. Finally, the computational complexity study confirmed the theoretical scaling predictions and showed that the proposed method achieves speedups of up to two orders of magnitude for the investigated dataset.

Overall, the results demonstrate that the fast mPOD preserves the interpretability and convergence properties of the classical formulation while drastically reducing computational cost. This opens the way to efficient multiscale analysis of large-scale datasets.

\subsection*{Acknowledgments}
{\small{This work was carried out as part of the Short Training Program (STP) of Marek Belda at the von Karman Institute. Marek Belda acknowledges the institutional support RVO:61388998 and the financial support provided by the Grant Agency of the Czech Technical University in Prague, grant No. SGS25/127/OHK2/3T/12. Martin Isoz acknowledges the financial support provided by the Ministry of Education, Youth, and Sports of the Czech Republic via the project No. CZ.02.01.01/00/23\_020/0008501 (METEX), co-funded by the European Union. M. A. Mendez is supported by the European Research Council (ERC, grant agreement No 101165479 RE-TWIST StG). Views and opinions expressed are however those of the authors only and do not necessarily reflect those of the European Union or the European Research Council. Neither the European Union nor the granting authority can be held responsible for them. L. Schena is supported by Fonds Wetenschappelijk Onderzoek (FWO), Proj. Number 1S67925N.
}}

\bibliographystyle{elsarticle-num-names}
\bibliography{references}

\clearpage
\appendix
\section{Results for block-structured matrices}
\label{sec:appendix}

This appendix collects the algebraic results used in Section~\ref{sec3} to justify the reduction of eigenvalue problems associated with block-sparse spectral correlation matrices.

\paragraph{Theorem A1}
Let
\begin{equation}
\label{eq:A0T}
\bm{A}=
\begin{bmatrix}
\bm{0} & \bm{0} & \bm{0} & \bm{0} & \bm{0}\\
\bm{0} & \bm{A}_1 & \bm{0} & \bm{A}_2 & \bm{0}\\
\bm{0} & \bm{0} & \bm{0} & \bm{0} & \bm{0}\\
\bm{0} & \bm{A}_3 & \bm{0} & \bm{A}_4 & \bm{0}\\
\bm{0} & \bm{0} & \bm{0} & \bm{0} & \bm{0}
\end{bmatrix}
\in\mathbb{C}^{N\times N},
\end{equation}
where the nonzero blocks do not overlap and are arranged so that
\[
\bm{A}_c=
\begin{bmatrix}
\bm{A}_1 & \bm{A}_2\\
\bm{A}_3 & \bm{A}_4
\end{bmatrix}
\in\mathbb{C}^{n\times n}
\]
is well defined. Then the nonzero eigenvalues of \(\bm{A}\) coincide with the eigenvalues of \(\bm{A}_c\).

\paragraph{Proof}
The characteristic polynomial of \(\bm{A}\) factorizes as
\begin{equation}
\det(\bm{A}-\lambda \bm{I})
=
\lambda^{\,N-n}\det(\bm{A}_c-\lambda \bm{I}),
\end{equation}
obtained by expanding the determinant along the zero-padded rows and columns. Hence, the spectrum of \(\bm{A}\) consists of the spectrum of \(\bm{A}_c\) together with \(N-n\) additional zero eigenvalues. \hfill\(\square\)

\paragraph{Theorem A2}
Let \(\lambda\neq 0\) be an eigenvalue of \(\bm{A}\) in \eqref{eq:A0T}, with associated eigenvector
\[
\bm{v}=
\begin{bmatrix}
\bm{v}_1\\ \bm{v}_2\\ \bm{v}_3\\ \bm{v}_4\\ \bm{v}_5
\end{bmatrix}.
\]
Then \(\bm{v}_1=\bm{v}_3=\bm{v}_5=\bm{0}\), and the nonzero part of \(\bm{v}\) is obtained from the eigenvector
\[
\bm{v}_c=
\begin{bmatrix}
\bm{v}_2\\ \bm{v}_4
\end{bmatrix}
\]
of \(\bm{A}_c\) associated with the same eigenvalue \(\lambda\).

\paragraph{Proof}
From \((\bm{A}-\lambda \bm{I}) \, \bm{v}=\bm{0}\), the rows corresponding to the zero blocks give
\[
-\lambda \, \bm{v}_1=\bm{0},\qquad
-\lambda \, \bm{v}_3=\bm{0},\qquad
-\lambda \, \bm{v}_5=\bm{0}.
\]
Since \(\lambda\neq 0\), it follows that
\[
\bm{v}_1=\bm{v}_3=\bm{v}_5=\bm{0}.
\]
The remaining equations reduce to
\begin{equation}
\begin{bmatrix}
\bm{A}_1-\lambda \bm{I} & \bm{A}_2\\
\bm{A}_3 & \bm{A}_4-\lambda \bm{I}
\end{bmatrix}
\begin{bmatrix}
\bm{v}_2\\ \bm{v}_4
\end{bmatrix}
=
(\bm{A}_c-\lambda \bm{I}) \, \bm{v}_c
=
\bm{0},
\end{equation}
so \(\bm{v}_c\) is an eigenvector of \(\bm{A}_c\), and \(\bm{v}\) is recovered by zero-padding. \hfill\(\square\)

\paragraph{Theorem A3}
Let \(\bm{A},\bm{B}\in\mathbb{C}^{N\times N}\) be two block-structured matrices of the form \eqref{eq:A0T}, with nonzero blocks supported on disjoint sets of rows and columns. Define
\[
\bm{C}=\bm{A}+\bm{B}.
\]
Then:
\begin{enumerate}
    \item every nonzero eigenpair of \(\bm{A}\) or \(\bm{B}\) is also an eigenpair of \(\bm{C}\);
    \item the nonzero spectrum of \(\bm{C}\) is the union of the nonzero spectra of \(\bm{A}\) and \(\bm{B}\).
\end{enumerate}

\paragraph{Proof}
Let \((\lambda,\bm{v})\) be a nonzero eigenpair of \(\bm{A}\). By Theorem~A2, the support of \(\bm{v}\) is confined to the rows and columns occupied by the nonzero blocks of \(\bm{A}\). Since \(\bm{B}\) is supported on a disjoint set of rows and columns, one has \(\bm{B} \, \bm{v}=\bm{0}\). Therefore
\[
\bm{C} \, \bm{v}
=
(\bm{A}+\bm{B})\, \bm{v}
=
\bm{A} \, \bm{v}
=
\lambda \, \bm{v}.
\]
The same argument holds for eigenpairs of \(\bm{B}\). Since the supports are disjoint, the nonzero invariant subspaces do not interact, and the nonzero spectrum of \(\bm{C}\) is exactly the union of the nonzero spectra of \(\bm{A}\) and \(\bm{B}\). \hfill\(\square\)
\end{document}